\documentclass[aps,prl,superscriptaddress,floatfix,twocolumn]{revtex4}

\usepackage{amsmath,amsfonts,amssymb,mathtools,braket,graphicx,enumitem}
\usepackage[utf8]{inputenc}
\usepackage[colorlinks=true,citecolor=blue,linkcolor=red,urlcolor=blue]{hyperref}
\usepackage[normalem]{ulem}
\usepackage[percent]{overpic}

\newcommand{\javr}[1]{\textcolor{red}{#1}}

\begin{document}

%\title{Multirefringence and anomalous caustics in multi-band systems}
\title{Snell's law in multirefringent systems}

\author{ J{\'o}zsef Cserti}
\author{Áron Holló}
\affiliation{Department of Physics of Complex Systems,
	ELTE E{\"o}tv{\"o}s Lor{\'a}nd University,
	H-1117 Budapest, P\'azm\'any P{\'e}ter s{\'e}t\'any 1/A, Hungary}

\author{László Oroszlány}
\affiliation{Department of Physics of Complex Systems,
	ELTE E{\"o}tv{\"o}s Lor{\'a}nd University,
	H-1117 Budapest, P\'azm\'any P{\'e}ter s{\'e}t\'any 1/A, Hungary\\
Wigner Research Centre for Physics, H-1525, Budapest, Hungary
}

\begin{abstract}

In anisotropic crystals, Maxwell's equations permit only birefringence for the propagation of light. 
Notwithstanding, multirefringent systems comprising more than two propagating modes exist, such as in electron optics and photonic crystals.
It has been demonstrated that in such systems, the propagation of waves in the short wavelength limit results in the formation of anomalous caustics. 
To calculate these caustic curves, we generalized Snell's law valid for reflection and refraction in multirefringent systems possessing more than one propagating mode.  
The emergence of anomalous caustics in the wave function patterns obtained from rigorous quantum mechanical calculations in electron-optical systems is confirmed. 
These calculations were performed on electron scattering from circularly gated potential regions applied to multilayer rhombohedral graphene. 
Our results may generate further work to explore more complex phenomena in multirefringent systems. 
 
\end{abstract}

\maketitle

%\section{Introduction}
%\label{intro:sec}
\textit{Introduction}.---
Birefringence, or double refraction
related to crystal anisotropy is a well-known phenomenon in optics~\cite{Born_Wolf_optics:book}. 
In the most general case, due to
Maxwell's equations, in anisotropic, non-magnetic, dielectric crystals, 
only birefringence is allowed. 
As it is well known, birefringence gives rise to a wide range of optical phenomena and several applications, see e.g.,~\cite{Hecht_optics:book}. 
Recently, bi-reflection of spin waves has been observed experimentally in magnonics system as well~\cite{hioki2020bi}.

Photonic crystals fabricated on the scale of the optical wavelength opened a new field in optics~\cite{PhysRevLett.90.077405}.  
For instance, in contrast to birefringence in dielectric crystals, multirefringence can be observed in trifringent photonic crystal waveguides~\cite{trirefringence_PRL.86.1526}. 
A theoretical study of trirefringence at the surface of metamaterials has also been reported recently~\cite{Diaz-Avino:16}.
The existence of triple refraction was proposed in nonlinear metamaterials as well~\cite{PhysRevA.86.013801}.

%Electron optics is another field where non-trivial wave propagation gives rise to exotic phenomena.  
%For instance, negative refractive indices, first theoretically proposed by Veselago and Pendry~\cite{Veselago_1968,Pendry_PhysRevLett.85.3966}
%have been experimentally realized in photonic crystals~\cite{Shalaev_1st_exp_negative_n_2005-ph,Shalaev_REVIEW_Nature_Photonics_2007,Cai2010} and in acoustic materials~\cite{sonic_bifreringence_Lu2007}.

Non-trivial wave propagation in materials with negative refractive indices gives rise to another exotic phenomenon. 
The existence of negative refraction indices has been theoretically proposed by Veselago and Pendry~\cite{Veselago_1968,Pendry_PhysRevLett.85.3966}
have been experimentally realized in photonic crystals~\cite{Shalaev_1st_exp_negative_n_2005-ph,Shalaev_REVIEW_Nature_Photonics_2007,Cai2010} and in acoustic materials~\cite{sonic_bifreringence_Lu2007}.
Research in this direction was initiated by the seminal work of Cheianov et al., proposing the realization of negative refractive index for electrons in graphene~\cite{Cheianov_Falko_neg_n_Science}. 
Electron focusing phenomenon in graphene has been further studied in Refs.~\cite{caustics_graphene_CsJ_PRL.99.246801,bilayer_electron_flow_CsJ_PRB.80.075416,Catastrophe_caustics_CsJ_PSSB_2010,bilayer_focusing_Peterfalvi_CsJ_2012,Dirac_rainbow_PhysRevB.90.235402,1_st_exp_GR_negative_n_Lee2015,Zhao_whispering_exp_science_aaa7469_2015,Electron_Mie_exp_Caridad2016,Jiang2017,PhysRevB.97.045413} and also in other two-dimensional (2D) electronic systems~\cite{Dirac_Weyl_PhysRevB.84.165115,Ulloa_PhysRevB.87.075420,perfect_caustics_PhysRevB.94.165405,Kekule_Naumis_PhysRevB.106.195413,PhysRevResearch.2.043245}. 
A perspective view of the realization of a 2D Dirac fermion-based microscope has now been considered~\cite{Boggild2017}.
In a recent publication, electron quantum optics in graphene has been reviewed~\cite{Chakraborti_Makk_Peter_review_2024}.
%~\cite{Makk_Peter_review:arxiv:1911.089144}.

In geometrical optics, classical ray approximation is the short-wavelength limit of electromagnetic theory.
The envelope of a ray family is called the caustic.
In the ray approximation, the density of the wave function patterns along the caustics is singular, thus they can be well seen even for finite wavelengths. 
Caustics have been studied extensively in the past (for a review see e.g. Berry and Upstill~\cite{Catastrophe_Optics_Berry_089} and Berry~\cite{sing_wave_caustics_Berry_105}).

\begin{figure}[htb]
\centering
\includegraphics[width=\columnwidth]{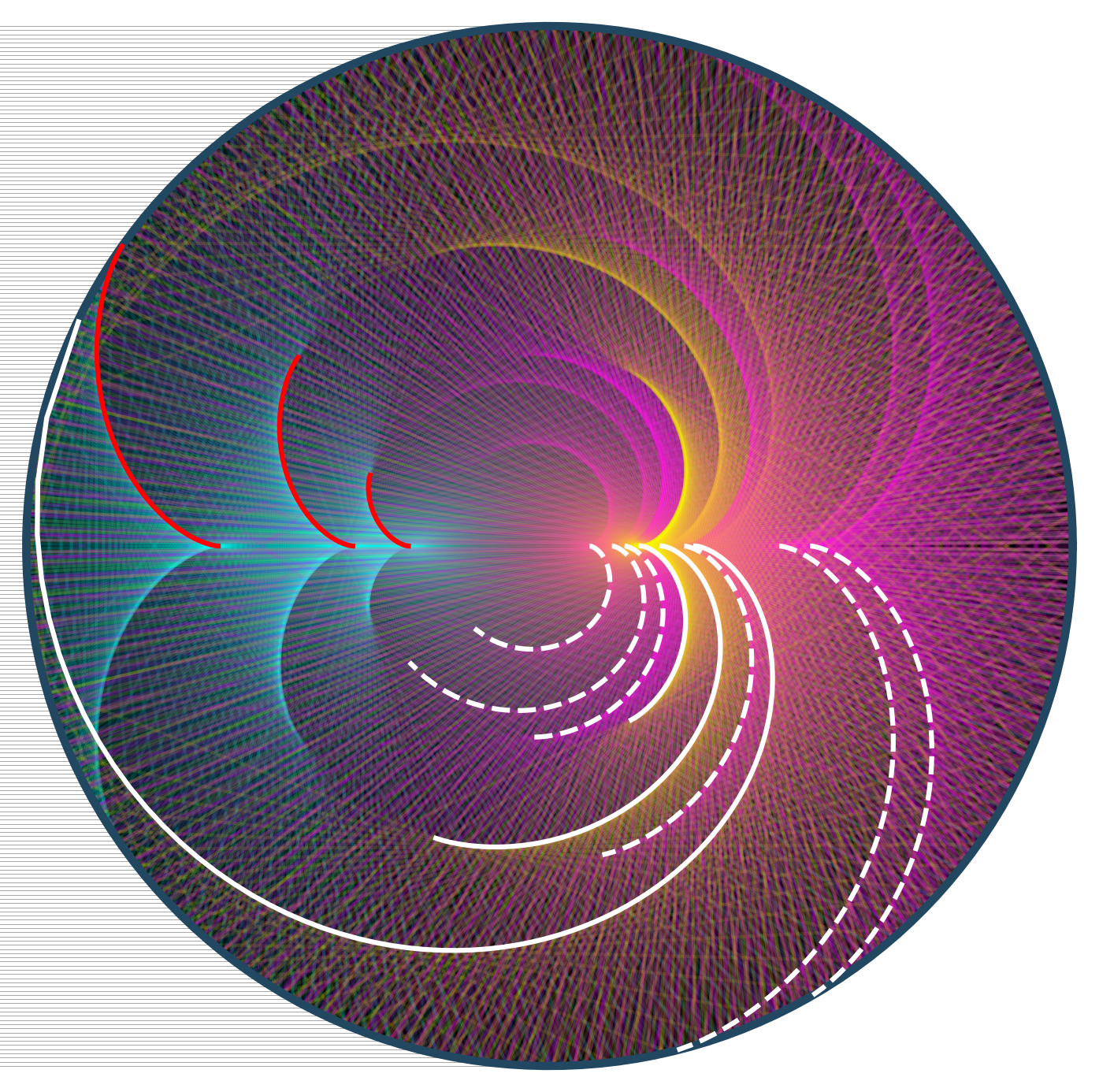}
\caption{Incident parallel rays from left (thin black solid lines) are refracted in three different directions (cyan-colored rays) at the circular surface, due to three different refractive indices of the region. The envelope of these rays, marked by red solid lines correspond to regular caustics.
These rays are reflected inside the circular region either without changing the refractive index (yellow-colored rays) or with an altered refractive index (magenta-colored rays). The envelopes of the yellow rays, marked by white solid lines are further regular caustics. The envelopes of rays with 
magenta color, highlighted by dashed white lines, are anomalous caustics. 
\label{geo_caustics_3:fig}}
\end{figure}

In this paper, we demonstrate that multirefringence, in contrast to birefringence as commonly observed in dielectric crystals, results in additional complexity. 
To this end, we consider the refraction of parallel rays incident on a circular region with more than two different refractive indices in ray approximation. 
As an illustration in Fig.~\ref{geo_caustics_3:fig} we depict characteristic features due to multirefringence in a region with three different refractive indices. 
Two types of caustic curves of the reflected rays can be formed inside the region interest:
(i) \textit{regular} caustics, where all rays are related to reflections with the same refractive index,  
(ii) \textit{anomalous} caustics (as we shall call them from now on), when the reflections inside the region involve more than one refractive index. 
The former is the usual caustics observed in a material with a single refractive index while the latter can only exist in multirefringent systems. 
From now on, we use the language of electron optics. However, our results can easily be extended to systems such as photonic crystals or acoustic waves in sonic crystals.

%We organize our manuscript as follows. First, we review concepts of geometrical optics underpinning the phenomena associated with multirefringence and anomalous caustics. 
%Second, we present a fully quantum mechanical scattering calculation, taking rhombohedral graphene as a template, underpinning the results obtained from geometrical optics. 
%Then, we conclude by summarising the general features of multirefringent systems.

%%% /Dropbox/AD/Multirefringence/elmelet/ray_in_out_cikk.ipynb
%%% /Dropbox/AD/Multirefringence/elmelet/make_fig_1.ipynb

\begin{figure}[hbt]
%%% .../AD/Multirefringence/elmelet/make_fig_1.ipynb
	\centering
	\includegraphics[width=\columnwidth]{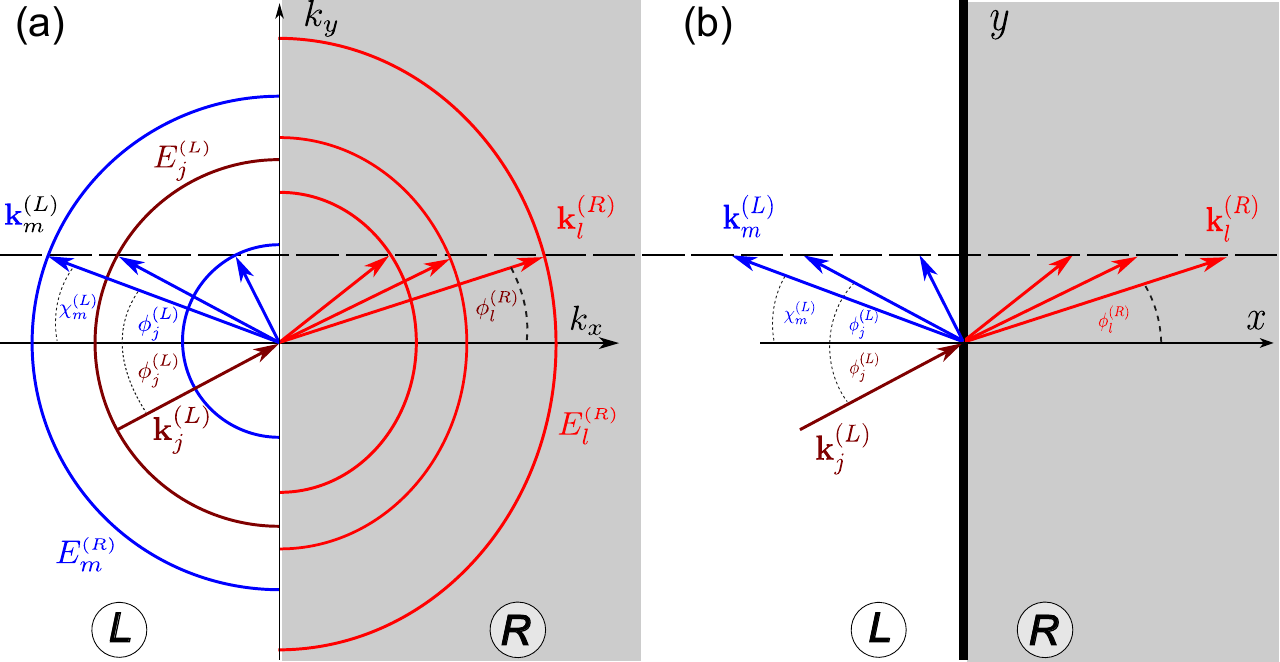}
        % /AD/Multirefringence/Cikk/figs/fig_1_disp_isotrope_pos_n.svg
	\caption{Multiple reflections and refractions for isotropic band structures. 
		(a) In $\mathbf{k}$-space the incident wave $\mathbf{k}_j^{(L)}$ (brown arrow) from the left side (L) of the interface 
		will be reflected (blue arrows) and refracted (red arrows) with wave vectors $\mathbf{k}_m^{(L)}$ 
		and $\mathbf{k}_l^{(R)}$ at the left and right sides of the interface (R), respectively. The constant energy surfaces (circles in 2D) are labeled by $j,m=1,2,\cdots,N_L$ and $l=1,2,\cdots,N_R$ at the left and right-hand sides of the interface, respectively.  For clarity in the figure, only three bands are shown on both sides. %The angles between the considered rays and the direction perpendicular to the interface are  denoted by $\phi_j^{(L/R)}$ and $\chi^{(L)}_m$, respectively.
		(b) shows the corresponding real-space trajectories of the reflected/refracted waves.
		\label{refl_geo_1:fig}}
\end{figure}

%\subsection{Multirefringence }
%\label{multi_ref:sub_sec}
\textit{Multirefringence}.---We now outline the characteristic features of multirefringence in multi-band systems. 
%For the sake of simplicity, only the two-dimensional case is presented here, since that can easily be generalized to the three-dimensional case.
Consider the propagation of free electrons through a flat interface in such systems. 
Assume that the number of plane wave solutions of the Schrödinger equation in the left $(L)$ and the right-hand $(R)$ sides of the interface with wave vector $\mathbf{k}$ are $N_L$ and $N_R$ ($N_{L/R} >1$) component spinor wave functions and the corresponding energy eigenvalues are $E^{(L/R)}_j(\mathbf{k})$, where $j,l=1,2,\cdots,N_{L/R}$. 
The system possesses $N_{L/R}$ number of bands at the two sides of the interface.
Here we assume that the system has isotropic band structures. 
In this case, on both sides of the interface, the group velocities $\mathbf{v}=\partial_{\mathbf{k}}E(\mathbf{k})/\hbar$ are parallel with the wave vector.
%In this case, the constant energy surfaces are circles, therefore the group velocities are proportional to the wave vector.
%%%% TODO put these ideas into the conclusions!!
%In this case, the constant energy surfaces are circles/spheres in two- and three-dimensions, respectively, therefore the group velocities are proportional to the wave vector. 
%The anisotropic case can be treated similarly.% However, the calculations are complicated since the direction of the group velocities (perpendicular to the constant energy surface), is not necessarily parallel to the wave vector.
To elucidate electron optical phenomena, one first needs to consider ray optics on a flat interface as depicted in Fig.~\ref{refl_geo_1:fig}.
For a given energy $E$ the allowed wave vectors $\mathbf{k}_j^{(L/R)}$ at the left and right side of the interface, respectively are the solutions 
of the equation $E^{(L/R)}_j(\mathbf{k}) = E$.
An incident ray with a wave vector $\mathbf{k}_j^{(L)}$ in general will be reflected/refracted into $N_{L/R}$ number of rays with corresponding wave vectors $\mathbf{k}_m^{(L/R)}$.
% (where $m=1,2,\cdots,N_{L,R}$).  
%Figure~\ref{refl_geo_1:fig} shows that foFurthermore, negative
%refractive indices, first theoretically proposed by Veselago and Pendry [4, 5] have been experimentally realized
%in photonic crystals [6–8] and in acoustic materials [9].
%Photonic crystals fabricated on the scale of the 
%two-dimensional multi-band systems and positive refractive index how the incident wave with wave vector $\mathbf{k}_j$  will be reflected $N_L$  and refracted $N_R$ number of different directions with wave vectors $\mathbf{k}_m^{(L)}$ (where $m=1,2,\cdots,N_L$) and $\mathbf{k}_l^{(R)}$ at the interface lying in the $y$ direction.

%\onecolumngrid

%\twocolumngrid
\textit{Generalized Snell's law}.---
For a given incident angle $\phi_j^{(L)}$ the reflection angles $\chi^{(L)}_m$ 
and the refracted angles $\phi^{(R)}_l$ can be determined from Snell's law which follows from the conservation of the $y$ components of the incident/reflected/refracted wave vectors at the flat interface:
\begin{subequations}
	\label{Snell_isotrope:eq}
	\begin{align}
		\label{Snell_isotrope_a:eq}
		k_j^{(L)}\sin \phi_j^{(L)} &= k_m^{(L)} \sin\chi^{(L)}_m, \quad  j,m=1,2,\dots,N_L \\[1ex]
		k_j^{(L)}\sin \phi_j^{(L)} &= k_l^{(R)} \sin\phi^{(R)}_l, \quad  l=1,2,\dots,N_R 
	\end{align}
\end{subequations}
where $k_{j,m}^{(L)} = \left|\mathbf{k}_{j,m}^{(L)}\right|$ and 
$k_{l}^{(R)} = \left|\mathbf{k}_{l}^{(R)}\right|$ are the magnitudes of the corresponding wave vectors.
These equations can be regarded as a generalization of Snell's laws valid for reflections and refractions in multirefringent systems.
Note that the same equations of Snell's law are valid for reflected and refracted elastic waves at the boundary between two different elastic media~\cite{Landau1986_VII}.
%Eq.~(\ref{Snell_isotrope:eq})

To demonstrate interband scattering highlighted above, we now consider one incoming plane wave (propagating from left to right as in Fig~\ref{geo_caustics_3:fig}.) outside the circular potential region.
For a given energy $E$, the sets of the allowed wave numbers inside and outside the potential 
region are denoted by
$\mathbb {K}^{<} = \left\{k^{<}_1,\dots, k^{<}_N\right\} $
and $\mathbb {K}^{>} = \left\{k^{>}_1,\dots, k^{>}_N \right\}$, respectively.
For simplicity, we assumed that the number of wave numbers inside and outside are the same $N$.
Naturally, the wave number $k_0$ of the incident wave is taken from the set  $\mathbb {K}^{>}$.
%Assume an incident plane wave with wave number $k_0 \in \mathbb {K}^{>} $. 
We now define a set of refractive indices $\hat{n}$ corresponding to the possible wave numbers inside the circular potential (the set $ \mathbb {K}^{<} $) as follows
\begin{align}
	\label{n_def:eq}
	\hat{n} &= \left\{ \left.\frac{k}{k_0} \right| \; k\in 
	\mathbb {K}^{<} \right\} \equiv 
	\left\{ n_1, \dots, n_N \right\}. 
\end{align}

Note that for anisotropic systems, the direction of the group velocities which is perpendicular to the constant energy surface, is not necessarily parallel to the wave vector.
%In isotropic systems,  the constant energy surfaces are circles/spheres in two- and three-dimensional, respectively, therefore the group velocities are proportional to the wave vector. 
In contrast to the dielectric crystals where only birefringence exists, in multi-band systems with $N > 2$ number of bands, multirefringence can arise.
Another intriguing feature of electron optics, realized for instance in graphene p--n junctions, is the feasibility of a negative refractive index.~\cite{1_st_exp_GR_negative_n_Lee2015,Zhao_whispering_exp_science_aaa7469_2015,Electron_Mie_exp_Caridad2016,Jiang2017}. In this case, we have $\mathbf{v}(\mathbf{k}) \mathbf{k} <0$.  
%We now discuss a few possible consequences of the multirefringence which does not arise in birefringent crystals.

%\subsection{Anomalous caustics }
%\label{anomal_caustics:sub_sec}
\textit{Anomalous caustics}.---To describe the formation of caustics in multi-band systems we consider a specific geometrical arrangement shown in Fig~\ref{geo_caustics_3:fig}. 
Inside the circular region, the refracted and the reflected rays of the incoming ray are straight lines called cords and numbered by  $p$ corresponding to $p-1$ reflections. 
The coordinates of the vertices $\mathbf{V}_i$ from which the rays are reflected at the circle of radius $R$, i.e., the endpoints of the cords can be parameterized as follows 
$\mathbf{V}_i = R \, {\left[\cos\varphi_i, \sin\varphi_i  \right]}^T$ for $i=0,1, \dots, p$, where $\varphi_0 \left(\alpha \right) =  \pi-\alpha$, and  
$\varphi_i \left(\alpha \right) = 
\pi-\alpha \pm \sum_{j=1}^{i} \bigl[ \pi - 2\chi_j\left(\alpha \right) \bigr]$ for $i=1,2, \dots, p $, 
and $\chi_j$ are the reflection angles at the vertex $\mathbf{V}_j$  which 
according to Eqs.~(\ref{Snell_isotrope:eq}) discussed above 
satisfy our generalized Snell's law
\begin{subequations}
	\label{Snell_refl_isotrope:eq}
	\begin{align}
		\sin \alpha &= n_1 \sin\chi_1, \\
		n_j \sin\chi_j &= n_{j+1} \sin\chi_{j+1} , 
		\quad \textrm{for} \quad j=1,2,\dots,p  ,
	\end{align}
\end{subequations}
where $\alpha$ is the incident angle of the incoming ray. 
The $+$ and $-$ signs in the definition of $\varphi_i \left(\alpha \right)$
correspond to the negative and positive refractive index, respectively, 
and the angles $\varphi_i$ are measured, as usual from the positive $x$ axis. 
However, in the absence of interband scattering, these angles are the same as they should be for a system with a single refraction index. 

\begin{figure}[hbt]
	\centering
	\includegraphics[scale=0.25]{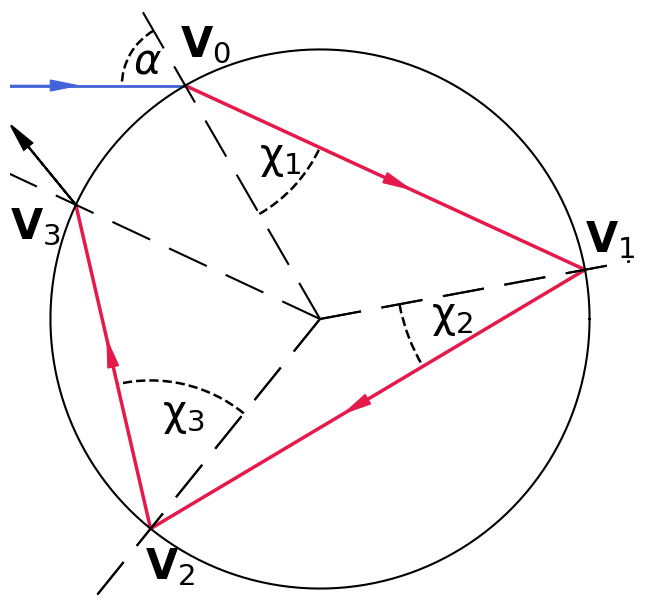}
	\put(-120,110){a)}
	\includegraphics[scale=0.25]{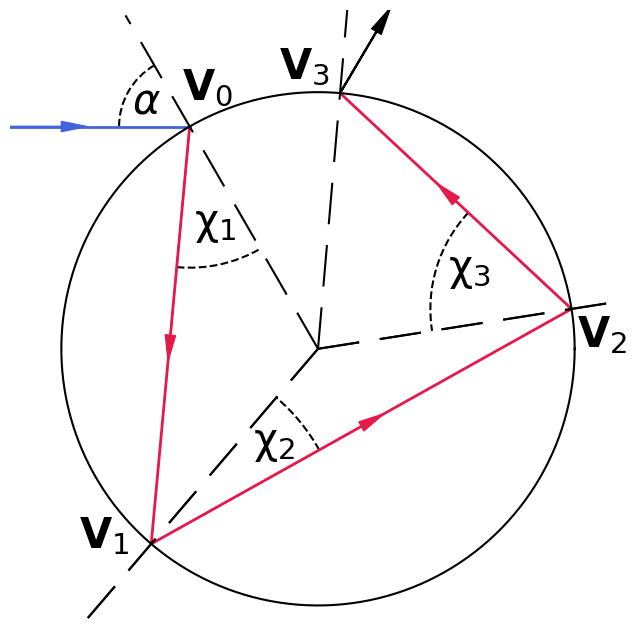}
	\put(-120,110){b)}
	\caption{Typical reflections of a ray inside the circular region for positive (a) and negative (b) refractive indices in case of $N=3$ number of bands and $p=3$ chords.  
		%	Here $\alpha = 60^{\circ}$, and inside the refractive indices are $n_1=1.5, n_2=2.5, n_3=1.1$ and outside $\bar{n}_2 = 2$. 
		\label{geo_2:fig}}
\end{figure}

Typical ray paths with reflection angles $\chi_j $ ($j= 1,2,3$) are shown in Fig.~\ref{geo_2:fig} for positive and negative refractive index. 
%%%  Dropbox/AD/Multirefringence/elmelet/ray_in_out_cikk.ipynb
%%% n1,n2,n3, n4 = (array([1.5, 2.5, 1.1, 2. ]), 2.0)
In this figure the reflection angles $\chi_j$ are different owing to 
the different wave numbers of the reflected rays. 
Therefore, each cord forms an anomalous caustic curve as the incident angle $\alpha$ varies and it can be calculated using differential geometry~\cite{stoker1989differential:book}.  
We find (see Supplemental Material~\cite{sajat_suppl_Mat_multirefringence:cikk})
the parametric equation of the regular and anomalous caustic curves corresponding to 
the $p$th chord as a function of $\alpha$: 
\begin{align}
	\label{rc:eq}
	{\mathbf{r}}_c^{(p)}(\alpha) & \equiv 
	\begin{bmatrix}
		x_c^{(p)}(\alpha) \\[1ex]
		y_c^{(p)}(\alpha)
	\end{bmatrix} =
	\frac{\mathbf{V}_{p-1}\, \dot{\varphi}_{p}  
		+ \mathbf{V}_{p}\, \dot{\varphi}_{p-1} }{\dot{\varphi}_{p-1}+\dot{\varphi}_p},
\end{align}
where the dot denotes the derivative with respect to $\alpha$. 
Thus, $\dot{\varphi}_i = \frac{\partial \varphi_i}{\partial \alpha} = 
-1 \mp 2 \sum_{j=1}^{i} \, \dot{\chi}_j$, and from the Snell's law~(\ref{Snell_refl_isotrope:eq}) we have 
$\dot{\chi}_j =\cos\alpha/\left(n_j \cos\chi_j\right)$.

To highlight the characteristic features of multirefringence discussed above, we present a full quantum mechanical solution of the scattering problem of electrons impinging on a circular potential region, shown in Fig.~\ref{geo_caustics_3:fig}. 
To this end, we carried out the calculations in rhombohedral graphene multilayers, widely studied in the literature~\cite{Guinea_stacks_PhysRevB.73.245426,Min_MacDonald_10.1143/PTPS.176.227,Katsnelson_ABC_PRB.84.125455,Van_Duppen_Peeters_2013,PhysRevB.77.045429}. 
The Hamiltonian of $N$-layer rhombohedral graphene has $N$ different positive energy electronic bands. Consequently, for a given energy there can, in general, be $N$ different wave numbers in one valley for states propagating in one direction. %Thus this $N$-layer ABC-stacked graphene could be an experimental realization of a multirefringent quantum electron-optical system.
The details of the quantum mechanical calculations of the scattering problem are given 
in the Supplemental Material~\cite{sajat_suppl_Mat_multirefringence:cikk}.

The calculations can be regarded as an extension of that carried out for single and bilayer graphene~\cite{caustics_graphene_CsJ_PRL.99.246801,bilayer_electron_flow_CsJ_PRB.80.075416,Catastrophe_caustics_CsJ_PSSB_2010,bilayer_focusing_Peterfalvi_CsJ_2012}.

\begin{figure*}[t]
	\centering
	\begin{tabular}{cc}
		\includegraphics[width=0.80\linewidth]{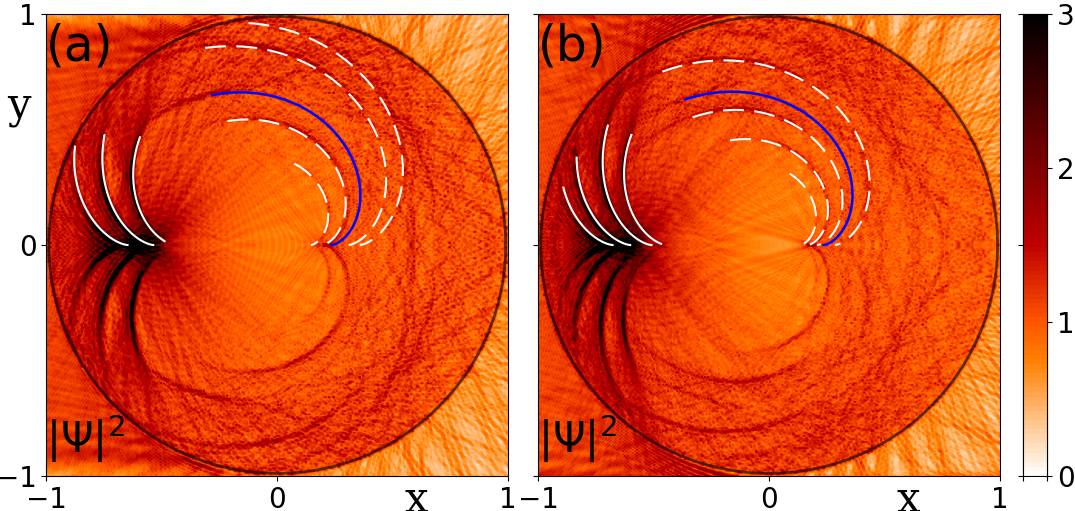}&
	\end{tabular}
	\caption{
		The intensity of the wave function inside the circularly gated potential for ABC graphene  (a) for $N=3$ layers (b) and $N=4$. 
		Using Eq.~(\ref{rc:eq}) for $p=1$ and $p=2$ the regular (white and blue solid lines, respectively),  
		and for $p=2$ the anomalous (white dashed lines) caustic curves are plotted. 
		The coordinates $x$ and $y$ are in units of $R$ and $R = 3000\, a$, where $a$ is the distance of two
		adjacent carbon atoms.
		The refractive indices of the circular regions for $N=3$ and $N=4$ are 
		$\hat{n} = \left\{-1.25, -0.86, -0.54 \right\}$ and 
		$\hat{n} = \left\{-1.24, -0.96, -0.66, -0.48 \right\}$, respectively.
		\label{QM_N_3_4:fig}}
\end{figure*}

As an example of multirefringence, in Fig.~\ref{QM_N_3_4:fig} we show the wave function pattern obtained from our fully quantum mechanical calculations for rhombohedral graphene with $N=3$ and $N=4$ layers, together with the caustics curves calculated from our analytic formula~(\ref{rc:eq}).
As can be seen, the caustic curves are well separated, and fit well to the intensity maxima of the calculated wave functions (for symmetry reasons only half of the caustic curves are shown to compare with the numerical results). Since caustics are related to geometrical optics, we cannot expect full agreement with the quantum mechanical result because the classical result does not take into account the quantum mechanical interference effect, and reflection and transmission amplitudes at the circular potential region.
Thus in the figure, we only highlight caustic curves that are well-visible in the wave function pattern.  
The scattered wave function pattern is calculated  in the semiclassical regime, i.e., in short wavelength limit, 
$k R \gg 1$ for incoming and refractive waves, where $R$ is the radius of the circular region of the scattering potential.
In general, it is clear that for the $p$th caustics (corresponding to the $p$th chord) in a multirefringent system with $N$ number of refraction indices, the number of regular caustics is $N$, and the number of anomalous caustics is $N^p-N$. This is well illustrated in Fig.~\ref{geo_caustics_3:fig}, where there are three regular caustics for all $p$, and for $p=2$ six anomalous caustic curves can be seen, in the case of $N=3$.

The locations of the cusps in the caustic curves,  
$(x,y) = (	x^{(p)}_{\mathrm{cusp}},0)$ can be derived from Eq.~(\ref{rc:eq}) in the limit of $\alpha \to 0$. 
For regular caustics, the result is the same as that obtained 
in Ref.~\onlinecite{caustics_graphene_CsJ_PRL.99.246801} for arbitrary $p$:
\begin{align}
	\label{cusp_anomal_p1:eq}
	x^{(p)}_{\mathrm{cusp}} = \frac{(-1)^p}{|n|-1+2p}R, 
\end{align}
where   $n$ is the refractive index of the circular region. This expression, for the location of the cusps, can be generalized to an arbitrary number of refractive indices encompassing regular and anomalous caustic curves.
As an example of two different refractive indices the locations of the two cusps with $p=2$ chords we obtained
\begin{align}
	\label{cusp_anomal_p2:eq}
	x^{(2)}_{\mathrm{cusp}} = \frac{|n_1|}{|n_1|+2|n_2|+|n_1| |n_2|}R, 
	\quad \mathrm{and} \quad \{n_1 \leftrightarrow n_2\}. 
\end{align}
Note that as a consequence of Snell's law (\ref{Snell_refl_isotrope:eq}) for reflection,  the positions of the cusps are not the same for the exchange of the two refractive indices. Similar expressions can be derived for $p>2$.

Using Eqs.~(\ref{cusp_anomal_p1:eq}) and~(\ref{cusp_anomal_p2:eq}) the corresponding cusp positions are listed in Table~\ref{merged_cusps:table} for $N=3$ and $N=4$. 
Note that in this table for regular (anomalous) caustics the refractive indices are the same (different).
One can see from the table that these cusps are well distinguishable.  
%\subsection{Anomalous rainbow }
%\label{rainbow:sub_sec}
\begin{table}[htb]
	\centering
	\begin{tabular}{|c|c||c|c||c|c|} 
		\hline
		\multicolumn{2}{|c||}{$x^{(p=1)}_{\mathrm{cusp}}$} & \multicolumn{2}{c||}{$x^{(p=2)}_{\mathrm{cusp}}$ (N=3)} & \multicolumn{2}{c|}{$x^{(p=2)}_{\mathrm{cusp}}$ (N=4)} \\ \hline
		$N=3$ & $N=4$ & $n_1 \rightarrow n_2$ & $x^{(p=2)}_{\mathrm{cusp}}$ & $n_1 \rightarrow n_2$ & $x^{(p=2)}_{\mathrm{cusp}}$ \\ \hline \hline
		-0.44 & -0.45 & {-0.54 $\rightarrow$ -1.25} & {0.15} & {-0.48 $\rightarrow$ -1.24} & {0.14} \\ \hline
		-0.54 & -0.51 & {-0.86 $\rightarrow$ -1.25} & {0.19} & {-0.66 $\rightarrow$ -1.24} & {0.17} \\ \hline
		-0.65 & -0.60 & -1.25 $\rightarrow$ -1.25 & 0.24 & {-0.96 $\rightarrow$ -1.24} & {0.21} \\ \hline
		-     & -0.68 & {-1.25 $\rightarrow$ -0.86} & {0.31} & -1.24 $\rightarrow$ -1.24 & 0.24 \\ \hline
		-     & -     & {-0.86 $\rightarrow$ -0.54} & {0.36} & {-1.24 $\rightarrow$ -0.96} & {0.29} \\ \hline
	\end{tabular}
	\caption{The locations of the cusps $x^{(p=1)}_{\mathrm{cusp}}$ for $p=1$ chords, and $x^{(p=2)}_{\mathrm{cusp}}$
 for $p=2$ chords of the caustic curves shown in Fig.~\ref{QM_N_3_4:fig}  in case of $N=3$ and $N=4$ layers.
 The locations are in units of $R$.
 For $p=1$ there are only regular caustic curves. 
 In the case of $p=2$ the two refractive indices $n_1$ and $n_2$ are the same for regular caustics, 
 while for different $n_1$ and $n_2$ the caustics are anomalous (red numbers).
 }
	\label{merged_cusps:table}
\end{table}

We would like to emphasize that the results for caustic curves, cusp locations, and their agreement with the quantum mechanical solution are in alignment with previous studies on bilayer graphene~\cite{bilayer_electron_flow_CsJ_PRB.80.075416}. We present further evidence of this in the Supplemental Material~\cite{sajat_suppl_Mat_multirefringence:cikk}.

%For graphene in the presence of Rashba spin-orbit interaction has recently been studied by Asmar and Ulloa~\cite{Ulloa_PhysRevB.87.075420}. 
%A theoretically similar two-band system, graphene in the presence of Rashba spin-orbit interaction, has recently been studied by Asmar and Ulloa~\cite{Ulloa_PhysRevB.87.075420}. This work highlighted the locations of the cusps associated with regular caustic curves. 
%Even though their quantum mechanical results clearly show the presence of anomalous caustic curves, the semiclassical analysis of these is omitted from their work. We performed the numerical calculations for their system (not shown here), and found that the anomalous caustics agree well with those calculated from Eq.~(\ref{rc:eq}). Thus, this example also underlines our theory of the anomalous caustics that arise in multirefringent systems.  

\textit{Conclusions}.---
We studied the wave propagation in materials with multiple propagating modes.
We generalized Snell's law for reflections and refractions in the short wavelength limit corresponding to the geometrical optics-like treatment.  
We demonstrated that interband scattering leads to the formation of  \textit{anomalous} caustics, which are absent in systems possessing only one propagating mode.
To explore the quantum-to-classical transition in electron-optical systems we calculated electron scattering from circular gated potential regions applied to multilayer rhombohedral graphene.
The anomalous caustics are well visible in the wave function patterns obtained from rigorous quantum mechanical calculations in short wavelength limits.
Our comprehensive study of multirefringence and anomalous caustics may inspire further works for exploring the phenomena related to multirefringent systems not only in electron optics 
but also in photonic crystals, acoustic materials, and magnonics.

%We also believe that the well-known Fresnel formulas valid for optics can be generalized to multirefringent systems. 

%\subsection{ABC graphene stack}
%\label{ABC:sec}

..........................

%\section*{Acknowledgments}
We thank M.~Berry, C.~W.~J.~Beenakker, C.~J.~Lambert and Gy.~D\'avid for valuable discussions.  
This research was supported by the Ministry of Culture and Innovation and the National Research,
Development and Innovation Office within the Quantum Information National Laboratory of Hungary
(Grant No. 2022-2.1.1-NL-2022-00004), National Research, Development and Innovation Office (NKFIH) through Grant Nos. K134437.	
This project is supported by the TRILMAX Horizon Europe consortium (Grant No. 101159646).

%\bibliographystyle{/home/cserti/Dropbox/programok/tex/prsty}
%\bibliography{cikkek_AB.bib}

\bibliographystyle{apsrev4-2}
%\bibliography{/home/cserti/Dropbox/AD/BibTeX/cikkek.bib}
\bibliography{cikkek.bib}

\begin{thebibliography}{41}%
\makeatletter
\providecommand \@ifxundefined [1]{%
 \@ifx{#1\undefined}
}%
\providecommand \@ifnum [1]{%
 \ifnum #1\expandafter \@firstoftwo
 \else \expandafter \@secondoftwo
 \fi
}%
\providecommand \@ifx [1]{%
 \ifx #1\expandafter \@firstoftwo
 \else \expandafter \@secondoftwo
 \fi
}%
\providecommand \natexlab [1]{#1}%
\providecommand \enquote  [1]{``#1''}%
\providecommand \bibnamefont  [1]{#1}%
\providecommand \bibfnamefont [1]{#1}%
\providecommand \citenamefont [1]{#1}%
\providecommand \href@noop [0]{\@secondoftwo}%
\providecommand \href [0]{\begingroup \@sanitize@url \@href}%
\providecommand \@href[1]{\@@startlink{#1}\@@href}%
\providecommand \@@href[1]{\endgroup#1\@@endlink}%
\providecommand \@sanitize@url [0]{\catcode `\\12\catcode `\$12\catcode
  `\&12\catcode `\#12\catcode `\^12\catcode `\_12\catcode `\%12\relax}%
\providecommand \@@startlink[1]{}%
\providecommand \@@endlink[0]{}%
\providecommand \url  [0]{\begingroup\@sanitize@url \@url }%
\providecommand \@url [1]{\endgroup\@href {#1}{\urlprefix }}%
\providecommand \urlprefix  [0]{URL }%
\providecommand \Eprint [0]{\href }%
\providecommand \doibase [0]{https://doi.org/}%
\providecommand \selectlanguage [0]{\@gobble}%
\providecommand \bibinfo  [0]{\@secondoftwo}%
\providecommand \bibfield  [0]{\@secondoftwo}%
\providecommand \translation [1]{[#1]}%
\providecommand \BibitemOpen [0]{}%
\providecommand \bibitemStop [0]{}%
\providecommand \bibitemNoStop [0]{.\EOS\space}%
\providecommand \EOS [0]{\spacefactor3000\relax}%
\providecommand \BibitemShut  [1]{\csname bibitem#1\endcsname}%
\let\auto@bib@innerbib\@empty
%</preamble>
\bibitem [{\citenamefont {Born}\ \emph {et~al.}(1999)\citenamefont {Born},
  \citenamefont {Wolf},\ and\ \citenamefont {Bhatia}}]{Born_Wolf_optics:book}%
  \BibitemOpen
  \bibfield  {author} {\bibinfo {author} {\bibfnamefont {M.}~\bibnamefont
  {Born}}, \bibinfo {author} {\bibfnamefont {E.}~\bibnamefont {Wolf}},\ and\
  \bibinfo {author} {\bibfnamefont {A.~B.}\ \bibnamefont {Bhatia}},\
  }\href@noop {} {\emph {\bibinfo {title} {Principles of Optics:
  Electromagnetic Theory of Propagation, Interference and Diffraction of
  Light}}},\ \bibinfo {edition} {7th}\ ed.\ (\bibinfo  {publisher} {Cambridge
  University Press},\ \bibinfo {year} {1999})\BibitemShut {NoStop}%
\bibitem [{\citenamefont {Hecht}(2002)}]{Hecht_optics:book}%
  \BibitemOpen
  \bibfield  {author} {\bibinfo {author} {\bibfnamefont {E.}~\bibnamefont
  {Hecht}},\ }\href@noop {} {\emph {\bibinfo {title} {Optics}}},\ \bibinfo
  {edition} {4th}\ ed.\ (\bibinfo  {publisher} {Addison-Wesley},\ \bibinfo
  {year} {2002})\BibitemShut {NoStop}%
\bibitem [{\citenamefont {Hioki}\ \emph {et~al.}(2020)\citenamefont {Hioki},
  \citenamefont {Hashimoto},\ and\ \citenamefont {Saitoh}}]{hioki2020bi}%
  \BibitemOpen
  \bibfield  {author} {\bibinfo {author} {\bibfnamefont {T.}~\bibnamefont
  {Hioki}}, \bibinfo {author} {\bibfnamefont {Y.}~\bibnamefont {Hashimoto}},\
  and\ \bibinfo {author} {\bibfnamefont {E.}~\bibnamefont {Saitoh}},\ }\href
  {https://doi.org/10.1038/s42005-020-00455-6} {\bibfield  {journal} {\bibinfo
  {journal} {Communications Physics}\ }\textbf {\bibinfo {volume} {3}},\
  \bibinfo {pages} {188} (\bibinfo {year} {2020})}\BibitemShut {NoStop}%
\bibitem [{\citenamefont {Smith}\ and\ \citenamefont
  {Schurig}(2003)}]{PhysRevLett.90.077405}%
  \BibitemOpen
  \bibfield  {author} {\bibinfo {author} {\bibfnamefont {D.~R.}\ \bibnamefont
  {Smith}}\ and\ \bibinfo {author} {\bibfnamefont {D.}~\bibnamefont
  {Schurig}},\ }\href {https://doi.org/10.1103/PhysRevLett.90.077405}
  {\bibfield  {journal} {\bibinfo  {journal} {Phys. Rev. Lett.}\ }\textbf
  {\bibinfo {volume} {90}},\ \bibinfo {pages} {077405} (\bibinfo {year}
  {2003})}\BibitemShut {NoStop}%
\bibitem [{\citenamefont {Netti}\ \emph {et~al.}(2001)\citenamefont {Netti},
  \citenamefont {Harris}, \citenamefont {Baumberg}, \citenamefont {Whittaker},
  \citenamefont {Charlton}, \citenamefont {Zoorob},\ and\ \citenamefont
  {Parker}}]{trirefringence_PRL.86.1526}%
  \BibitemOpen
  \bibfield  {author} {\bibinfo {author} {\bibfnamefont {M.~C.}\ \bibnamefont
  {Netti}}, \bibinfo {author} {\bibfnamefont {A.}~\bibnamefont {Harris}},
  \bibinfo {author} {\bibfnamefont {J.~J.}\ \bibnamefont {Baumberg}}, \bibinfo
  {author} {\bibfnamefont {D.~M.}\ \bibnamefont {Whittaker}}, \bibinfo {author}
  {\bibfnamefont {M.~B.~D.}\ \bibnamefont {Charlton}}, \bibinfo {author}
  {\bibfnamefont {M.~E.}\ \bibnamefont {Zoorob}},\ and\ \bibinfo {author}
  {\bibfnamefont {G.~J.}\ \bibnamefont {Parker}},\ }\href
  {https://doi.org/10.1103/PhysRevLett.86.1526} {\bibfield  {journal} {\bibinfo
   {journal} {Phys. Rev. Lett.}\ }\textbf {\bibinfo {volume} {86}},\ \bibinfo
  {pages} {1526} (\bibinfo {year} {2001})}\BibitemShut {NoStop}%
\bibitem [{\citenamefont {D\'{i}az-Avi\~{n}\'{o}}\ \emph
  {et~al.}(2016)\citenamefont {D\'{i}az-Avi\~{n}\'{o}}, \citenamefont {Pastor},
  \citenamefont {Zapata-Rodr\'{i}guez}, \citenamefont {Naserpour},
  \citenamefont {Koty\'{n}ski},\ and\ \citenamefont {Miret}}]{Diaz-Avino:16}%
  \BibitemOpen
  \bibfield  {author} {\bibinfo {author} {\bibfnamefont {C.}~\bibnamefont
  {D\'{i}az-Avi\~{n}\'{o}}}, \bibinfo {author} {\bibfnamefont {D.}~\bibnamefont
  {Pastor}}, \bibinfo {author} {\bibfnamefont {C.~J.}\ \bibnamefont
  {Zapata-Rodr\'{i}guez}}, \bibinfo {author} {\bibfnamefont {M.}~\bibnamefont
  {Naserpour}}, \bibinfo {author} {\bibfnamefont {R.}~\bibnamefont
  {Koty\'{n}ski}},\ and\ \bibinfo {author} {\bibfnamefont {J.~J.}\ \bibnamefont
  {Miret}},\ }\href {https://doi.org/10.1364/JOSAB.33.000116} {\bibfield
  {journal} {\bibinfo  {journal} {J. Opt. Soc. Am. B}\ }\textbf {\bibinfo
  {volume} {33}},\ \bibinfo {pages} {116} (\bibinfo {year} {2016})}\BibitemShut
  {NoStop}%
\bibitem [{\citenamefont {De~Lorenci}\ and\ \citenamefont
  {Pereira}(2012)}]{PhysRevA.86.013801}%
  \BibitemOpen
  \bibfield  {author} {\bibinfo {author} {\bibfnamefont {V.~A.}\ \bibnamefont
  {De~Lorenci}}\ and\ \bibinfo {author} {\bibfnamefont {J.~P.}\ \bibnamefont
  {Pereira}},\ }\href {https://doi.org/10.1103/PhysRevA.86.013801} {\bibfield
  {journal} {\bibinfo  {journal} {Phys. Rev. A}\ }\textbf {\bibinfo {volume}
  {86}},\ \bibinfo {pages} {013801} (\bibinfo {year} {2012})}\BibitemShut
  {NoStop}%
\bibitem [{\citenamefont {Veselago}(1968)}]{Veselago_1968}%
  \BibitemOpen
  \bibfield  {author} {\bibinfo {author} {\bibfnamefont {V.~G.}\ \bibnamefont
  {Veselago}},\ }\href {https://doi.org/10.1070/PU1968v010n04ABEH003699}
  {\bibfield  {journal} {\bibinfo  {journal} {Soviet Physics Uspekhi}\ }\textbf
  {\bibinfo {volume} {10}},\ \bibinfo {pages} {509} (\bibinfo {year}
  {1968})}\BibitemShut {NoStop}%
\bibitem [{\citenamefont {Pendry}(2000)}]{Pendry_PhysRevLett.85.3966}%
  \BibitemOpen
  \bibfield  {author} {\bibinfo {author} {\bibfnamefont {J.~B.}\ \bibnamefont
  {Pendry}},\ }\href {https://link.aps.org/doi/10.1103/PhysRevLett.85.3966}
  {\bibfield  {journal} {\bibinfo  {journal} {Phys. Rev. Lett.}\ }\textbf
  {\bibinfo {volume} {85}},\ \bibinfo {pages} {3966} (\bibinfo {year}
  {2000})}\BibitemShut {NoStop}%
\bibitem [{\citenamefont {Shalaev}\ \emph {et~al.}(2005)\citenamefont
  {Shalaev}, \citenamefont {Cai}, \citenamefont {Chettiar}, \citenamefont
  {Yuan}, \citenamefont {Sarychev}, \citenamefont {Drachev},\ and\
  \citenamefont {Kildishev}}]{Shalaev_1st_exp_negative_n_2005-ph}%
  \BibitemOpen
  \bibfield  {author} {\bibinfo {author} {\bibfnamefont {V.~M.}\ \bibnamefont
  {Shalaev}}, \bibinfo {author} {\bibfnamefont {W.}~\bibnamefont {Cai}},
  \bibinfo {author} {\bibfnamefont {U.~K.}\ \bibnamefont {Chettiar}}, \bibinfo
  {author} {\bibfnamefont {H.-K.}\ \bibnamefont {Yuan}}, \bibinfo {author}
  {\bibfnamefont {A.~K.}\ \bibnamefont {Sarychev}}, \bibinfo {author}
  {\bibfnamefont {V.~P.}\ \bibnamefont {Drachev}},\ and\ \bibinfo {author}
  {\bibfnamefont {A.~V.}\ \bibnamefont {Kildishev}},\ }\href
  {https://doi.org/10.1364/OL.30.003356} {\bibfield  {journal} {\bibinfo
  {journal} {Opt Lett}\ }\textbf {\bibinfo {volume} {30}},\ \bibinfo {pages}
  {3356} (\bibinfo {year} {2005})}\BibitemShut {NoStop}%
\bibitem [{\citenamefont
  {Shalaev}(2007)}]{Shalaev_REVIEW_Nature_Photonics_2007}%
  \BibitemOpen
  \bibfield  {author} {\bibinfo {author} {\bibfnamefont {V.~M.}\ \bibnamefont
  {Shalaev}},\ }\href {https://doi.org/10.1038/nphoton.2006.49} {\bibfield
  {journal} {\bibinfo  {journal} {Nature Photonics}\ }\textbf {\bibinfo
  {volume} {1}},\ \bibinfo {pages} {41} (\bibinfo {year} {2007})}\BibitemShut
  {NoStop}%
\bibitem [{\citenamefont {Cai}\ and\ \citenamefont {Shalaev}(2010)}]{Cai2010}%
  \BibitemOpen
  \bibfield  {author} {\bibinfo {author} {\bibfnamefont {W.}~\bibnamefont
  {Cai}}\ and\ \bibinfo {author} {\bibfnamefont {V.}~\bibnamefont {Shalaev}},\
  }\href@noop {} {\emph {\bibinfo {title} {Optical Metamaterials: Fundamentals
  and Applications}}}\ (\bibinfo  {publisher} {Springer},\ \bibinfo {address}
  {New York},\ \bibinfo {year} {2010})\BibitemShut {NoStop}%
\bibitem [{\citenamefont {Lu}\ \emph {et~al.}(2007)\citenamefont {Lu},
  \citenamefont {Zhang}, \citenamefont {Feng}, \citenamefont {Zhao},
  \citenamefont {Chen}, \citenamefont {Mao}, \citenamefont {Zi}, \citenamefont
  {Zhu}, \citenamefont {Zhu},\ and\ \citenamefont
  {Ming}}]{sonic_bifreringence_Lu2007}%
  \BibitemOpen
  \bibfield  {author} {\bibinfo {author} {\bibfnamefont {M.-H.}\ \bibnamefont
  {Lu}}, \bibinfo {author} {\bibfnamefont {C.}~\bibnamefont {Zhang}}, \bibinfo
  {author} {\bibfnamefont {L.}~\bibnamefont {Feng}}, \bibinfo {author}
  {\bibfnamefont {J.}~\bibnamefont {Zhao}}, \bibinfo {author} {\bibfnamefont
  {Y.-F.}\ \bibnamefont {Chen}}, \bibinfo {author} {\bibfnamefont {Y.-W.}\
  \bibnamefont {Mao}}, \bibinfo {author} {\bibfnamefont {J.}~\bibnamefont
  {Zi}}, \bibinfo {author} {\bibfnamefont {Y.-Y.}\ \bibnamefont {Zhu}},
  \bibinfo {author} {\bibfnamefont {S.-N.}\ \bibnamefont {Zhu}},\ and\ \bibinfo
  {author} {\bibfnamefont {N.-B.}\ \bibnamefont {Ming}},\ }\href
  {https://doi.org/10.1038/nmat1987} {\bibfield  {journal} {\bibinfo  {journal}
  {Nature Materials}\ }\textbf {\bibinfo {volume} {6}},\ \bibinfo {pages} {744}
  (\bibinfo {year} {2007})}\BibitemShut {NoStop}%
\bibitem [{\citenamefont {Cheianov}\ \emph {et~al.}(2007)\citenamefont
  {Cheianov}, \citenamefont {Fal'ko},\ and\ \citenamefont
  {Altshuler}}]{Cheianov_Falko_neg_n_Science}%
  \BibitemOpen
  \bibfield  {author} {\bibinfo {author} {\bibfnamefont {V.~V.}\ \bibnamefont
  {Cheianov}}, \bibinfo {author} {\bibfnamefont {V.}~\bibnamefont {Fal'ko}},\
  and\ \bibinfo {author} {\bibfnamefont {B.~L.}\ \bibnamefont {Altshuler}},\
  }\href {https://doi.org/10.1126/science.1138020} {\bibfield  {journal}
  {\bibinfo  {journal} {Science}\ }\textbf {\bibinfo {volume} {315}},\ \bibinfo
  {pages} {1252} (\bibinfo {year} {2007})}\BibitemShut {NoStop}%
\bibitem [{\citenamefont {Cserti}\ \emph {et~al.}(2007)\citenamefont {Cserti},
  \citenamefont {P\'alyi},\ and\ \citenamefont
  {P\'eterfalvi}}]{caustics_graphene_CsJ_PRL.99.246801}%
  \BibitemOpen
  \bibfield  {author} {\bibinfo {author} {\bibfnamefont {J.}~\bibnamefont
  {Cserti}}, \bibinfo {author} {\bibfnamefont {A.}~\bibnamefont {P\'alyi}},\
  and\ \bibinfo {author} {\bibfnamefont {C.}~\bibnamefont {P\'eterfalvi}},\
  }\href {https://doi.org/10.1103/PhysRevLett.99.246801} {\bibfield  {journal}
  {\bibinfo  {journal} {Phys. Rev. Lett.}\ }\textbf {\bibinfo {volume} {99}},\
  \bibinfo {pages} {246801} (\bibinfo {year} {2007})}\BibitemShut {NoStop}%
\bibitem [{\citenamefont {P\'eterfalvi}\ \emph {et~al.}(2009)\citenamefont
  {P\'eterfalvi}, \citenamefont {P\'alyi},\ and\ \citenamefont
  {Cserti}}]{bilayer_electron_flow_CsJ_PRB.80.075416}%
  \BibitemOpen
  \bibfield  {author} {\bibinfo {author} {\bibfnamefont {C.}~\bibnamefont
  {P\'eterfalvi}}, \bibinfo {author} {\bibfnamefont {A.}~\bibnamefont
  {P\'alyi}},\ and\ \bibinfo {author} {\bibfnamefont {J.}~\bibnamefont
  {Cserti}},\ }\href {https://doi.org/10.1103/PhysRevB.80.075416} {\bibfield
  {journal} {\bibinfo  {journal} {Phys. Rev. B}\ }\textbf {\bibinfo {volume}
  {80}},\ \bibinfo {pages} {075416} (\bibinfo {year} {2009})}\BibitemShut
  {NoStop}%
\bibitem [{\citenamefont {Péterfalvi}\ \emph {et~al.}(2010)\citenamefont
  {Péterfalvi}, \citenamefont {Pályi}, \citenamefont {Rusznyák},
  \citenamefont {Koltai},\ and\ \citenamefont
  {Cserti}}]{Catastrophe_caustics_CsJ_PSSB_2010}%
  \BibitemOpen
  \bibfield  {author} {\bibinfo {author} {\bibfnamefont {C.}~\bibnamefont
  {Péterfalvi}}, \bibinfo {author} {\bibfnamefont {A.}~\bibnamefont {Pályi}},
  \bibinfo {author} {\bibfnamefont {{\'A}.}~\bibnamefont {Rusznyák}}, \bibinfo
  {author} {\bibfnamefont {J.}~\bibnamefont {Koltai}},\ and\ \bibinfo {author}
  {\bibfnamefont {J.}~\bibnamefont {Cserti}},\ }\href
  {https://doi.org/10.1002/pssb.201000160} {\bibfield  {journal} {\bibinfo
  {journal} {Phys. Status Solidi B}\ }\textbf {\bibinfo {volume} {247}},\
  \bibinfo {pages} {2949} (\bibinfo {year} {2010})}\BibitemShut {NoStop}%
\bibitem [{\citenamefont {Péterfalvi}\ \emph {et~al.}(2012)\citenamefont
  {Péterfalvi}, \citenamefont {Oroszlány}, \citenamefont {Lambert},\ and\
  \citenamefont {Cserti}}]{bilayer_focusing_Peterfalvi_CsJ_2012}%
  \BibitemOpen
  \bibfield  {author} {\bibinfo {author} {\bibfnamefont {C.~G.}\ \bibnamefont
  {Péterfalvi}}, \bibinfo {author} {\bibfnamefont {L.}~\bibnamefont
  {Oroszlány}}, \bibinfo {author} {\bibfnamefont {C.~J.}\ \bibnamefont
  {Lambert}},\ and\ \bibinfo {author} {\bibfnamefont {J.}~\bibnamefont
  {Cserti}},\ }\href {https://doi.org/10.1088/1367-2630/14/6/063028} {\bibfield
   {journal} {\bibinfo  {journal} {New Journal of Physics}\ }\textbf {\bibinfo
  {volume} {14}},\ \bibinfo {pages} {063028} (\bibinfo {year}
  {2012})}\BibitemShut {NoStop}%
\bibitem [{\citenamefont {Wu}\ and\ \citenamefont
  {Fogler}(2014)}]{Dirac_rainbow_PhysRevB.90.235402}%
  \BibitemOpen
  \bibfield  {author} {\bibinfo {author} {\bibfnamefont {J.-S.}\ \bibnamefont
  {Wu}}\ and\ \bibinfo {author} {\bibfnamefont {M.~M.}\ \bibnamefont
  {Fogler}},\ }\href {https://doi.org/10.1103/PhysRevB.90.235402} {\bibfield
  {journal} {\bibinfo  {journal} {Phys. Rev. B}\ }\textbf {\bibinfo {volume}
  {90}},\ \bibinfo {pages} {235402} (\bibinfo {year} {2014})}\BibitemShut
  {NoStop}%
\bibitem [{\citenamefont {Lee}\ \emph {et~al.}(2015)\citenamefont {Lee},
  \citenamefont {Park},\ and\ \citenamefont
  {Lee}}]{1_st_exp_GR_negative_n_Lee2015}%
  \BibitemOpen
  \bibfield  {author} {\bibinfo {author} {\bibfnamefont {G.-H.}\ \bibnamefont
  {Lee}}, \bibinfo {author} {\bibfnamefont {G.-H.}\ \bibnamefont {Park}},\ and\
  \bibinfo {author} {\bibfnamefont {H.-J.}\ \bibnamefont {Lee}},\ }\href
  {https://doi.org/10.1038/nphys3460} {\bibfield  {journal} {\bibinfo
  {journal} {Nature Physics}\ }\textbf {\bibinfo {volume} {11}},\ \bibinfo
  {pages} {925} (\bibinfo {year} {2015})}\BibitemShut {NoStop}%
\bibitem [{\citenamefont {Zhao}\ \emph {et~al.}(2015)\citenamefont {Zhao},
  \citenamefont {Wyrick}, \citenamefont {Natterer}, \citenamefont
  {Rodriguez-Nieva}, \citenamefont {Lewandowski}, \citenamefont {Watanabe},
  \citenamefont {Taniguchi}, \citenamefont {Levitov}, \citenamefont
  {Zhitenev},\ and\ \citenamefont
  {Stroscio}}]{Zhao_whispering_exp_science_aaa7469_2015}%
  \BibitemOpen
  \bibfield  {author} {\bibinfo {author} {\bibfnamefont {Y.}~\bibnamefont
  {Zhao}}, \bibinfo {author} {\bibfnamefont {J.}~\bibnamefont {Wyrick}},
  \bibinfo {author} {\bibfnamefont {F.~D.}\ \bibnamefont {Natterer}}, \bibinfo
  {author} {\bibfnamefont {J.~F.}\ \bibnamefont {Rodriguez-Nieva}}, \bibinfo
  {author} {\bibfnamefont {C.}~\bibnamefont {Lewandowski}}, \bibinfo {author}
  {\bibfnamefont {K.}~\bibnamefont {Watanabe}}, \bibinfo {author}
  {\bibfnamefont {T.}~\bibnamefont {Taniguchi}}, \bibinfo {author}
  {\bibfnamefont {L.~S.}\ \bibnamefont {Levitov}}, \bibinfo {author}
  {\bibfnamefont {N.~B.}\ \bibnamefont {Zhitenev}},\ and\ \bibinfo {author}
  {\bibfnamefont {J.~A.}\ \bibnamefont {Stroscio}},\ }\href
  {https://doi.org/10.1126/science.aaa7469} {\bibfield  {journal} {\bibinfo
  {journal} {Science}\ }\textbf {\bibinfo {volume} {348}},\ \bibinfo {pages}
  {672} (\bibinfo {year} {2015})}\BibitemShut {NoStop}%
\bibitem [{\citenamefont {Caridad}\ \emph {et~al.}(2016)\citenamefont
  {Caridad}, \citenamefont {Connaughton}, \citenamefont {Ott}, \citenamefont
  {Weber},\ and\ \citenamefont {Krsti{\'{c}}}}]{Electron_Mie_exp_Caridad2016}%
  \BibitemOpen
  \bibfield  {author} {\bibinfo {author} {\bibfnamefont {J.~M.}\ \bibnamefont
  {Caridad}}, \bibinfo {author} {\bibfnamefont {S.}~\bibnamefont
  {Connaughton}}, \bibinfo {author} {\bibfnamefont {C.}~\bibnamefont {Ott}},
  \bibinfo {author} {\bibfnamefont {H.~B.}\ \bibnamefont {Weber}},\ and\
  \bibinfo {author} {\bibfnamefont {V.}~\bibnamefont {Krsti{\'{c}}}},\ }\href
  {https://doi.org/10.1038/ncomms12894} {\bibfield  {journal} {\bibinfo
  {journal} {Nature Communications}\ }\textbf {\bibinfo {volume} {7}},\
  \bibinfo {pages} {12894} (\bibinfo {year} {2016})}\BibitemShut {NoStop}%
\bibitem [{\citenamefont {Jiang}\ \emph {et~al.}(2017)\citenamefont {Jiang},
  \citenamefont {Mao}, \citenamefont {Moldovan}, \citenamefont {Masir},
  \citenamefont {Li}, \citenamefont {Watanabe}, \citenamefont {Taniguchi},
  \citenamefont {Peeters},\ and\ \citenamefont {Andrei}}]{Jiang2017}%
  \BibitemOpen
  \bibfield  {author} {\bibinfo {author} {\bibfnamefont {Y.}~\bibnamefont
  {Jiang}}, \bibinfo {author} {\bibfnamefont {J.}~\bibnamefont {Mao}}, \bibinfo
  {author} {\bibfnamefont {D.}~\bibnamefont {Moldovan}}, \bibinfo {author}
  {\bibfnamefont {M.~R.}\ \bibnamefont {Masir}}, \bibinfo {author}
  {\bibfnamefont {G.}~\bibnamefont {Li}}, \bibinfo {author} {\bibfnamefont
  {K.}~\bibnamefont {Watanabe}}, \bibinfo {author} {\bibfnamefont
  {T.}~\bibnamefont {Taniguchi}}, \bibinfo {author} {\bibfnamefont {F.~M.}\
  \bibnamefont {Peeters}},\ and\ \bibinfo {author} {\bibfnamefont {E.~Y.}\
  \bibnamefont {Andrei}},\ }\href {https://doi.org/10.1038/nnano.2017.181}
  {\bibfield  {journal} {\bibinfo  {journal} {Nature Nanotechnology}\ }\textbf
  {\bibinfo {volume} {12}},\ \bibinfo {pages} {1045} (\bibinfo {year}
  {2017})}\BibitemShut {NoStop}%
\bibitem [{\citenamefont {Bai}\ \emph {et~al.}(2018)\citenamefont {Bai},
  \citenamefont {Zhou}, \citenamefont {Wei}, \citenamefont {Qiao},
  \citenamefont {Liu}, \citenamefont {Liu}, \citenamefont {Jiang},\ and\
  \citenamefont {He}}]{PhysRevB.97.045413}%
  \BibitemOpen
  \bibfield  {author} {\bibinfo {author} {\bibfnamefont {K.-K.}\ \bibnamefont
  {Bai}}, \bibinfo {author} {\bibfnamefont {J.-J.}\ \bibnamefont {Zhou}},
  \bibinfo {author} {\bibfnamefont {Y.-C.}\ \bibnamefont {Wei}}, \bibinfo
  {author} {\bibfnamefont {J.-B.}\ \bibnamefont {Qiao}}, \bibinfo {author}
  {\bibfnamefont {Y.-W.}\ \bibnamefont {Liu}}, \bibinfo {author} {\bibfnamefont
  {H.-W.}\ \bibnamefont {Liu}}, \bibinfo {author} {\bibfnamefont
  {H.}~\bibnamefont {Jiang}},\ and\ \bibinfo {author} {\bibfnamefont
  {L.}~\bibnamefont {He}},\ }\href {https://doi.org/10.1103/PhysRevB.97.045413}
  {\bibfield  {journal} {\bibinfo  {journal} {Phys. Rev. B}\ }\textbf {\bibinfo
  {volume} {97}},\ \bibinfo {pages} {045413} (\bibinfo {year}
  {2018})}\BibitemShut {NoStop}%
\bibitem [{\citenamefont {Lan}\ \emph {et~al.}(2011)\citenamefont {Lan},
  \citenamefont {Goldman}, \citenamefont {Bermudez}, \citenamefont {Lu},\ and\
  \citenamefont {\"Ohberg}}]{Dirac_Weyl_PhysRevB.84.165115}%
  \BibitemOpen
  \bibfield  {author} {\bibinfo {author} {\bibfnamefont {Z.}~\bibnamefont
  {Lan}}, \bibinfo {author} {\bibfnamefont {N.}~\bibnamefont {Goldman}},
  \bibinfo {author} {\bibfnamefont {A.}~\bibnamefont {Bermudez}}, \bibinfo
  {author} {\bibfnamefont {W.}~\bibnamefont {Lu}},\ and\ \bibinfo {author}
  {\bibfnamefont {P.}~\bibnamefont {\"Ohberg}},\ }\href
  {https://doi.org/10.1103/PhysRevB.84.165115} {\bibfield  {journal} {\bibinfo
  {journal} {Phys. Rev. B}\ }\textbf {\bibinfo {volume} {84}},\ \bibinfo
  {pages} {165115} (\bibinfo {year} {2011})}\BibitemShut {NoStop}%
\bibitem [{\citenamefont {Asmar}\ and\ \citenamefont
  {Ulloa}(2013)}]{Ulloa_PhysRevB.87.075420}%
  \BibitemOpen
  \bibfield  {author} {\bibinfo {author} {\bibfnamefont {M.~M.}\ \bibnamefont
  {Asmar}}\ and\ \bibinfo {author} {\bibfnamefont {S.~E.}\ \bibnamefont
  {Ulloa}},\ }\href {https://doi.org/10.1103/PhysRevB.87.075420} {\bibfield
  {journal} {\bibinfo  {journal} {Phys. Rev. B}\ }\textbf {\bibinfo {volume}
  {87}},\ \bibinfo {pages} {075420} (\bibinfo {year} {2013})}\BibitemShut
  {NoStop}%
\bibitem [{\citenamefont {Xu}\ and\ \citenamefont
  {Lai}(2016)}]{perfect_caustics_PhysRevB.94.165405}%
  \BibitemOpen
  \bibfield  {author} {\bibinfo {author} {\bibfnamefont {H.-Y.}\ \bibnamefont
  {Xu}}\ and\ \bibinfo {author} {\bibfnamefont {Y.-C.}\ \bibnamefont {Lai}},\
  }\href {https://doi.org/10.1103/PhysRevB.94.165405} {\bibfield  {journal}
  {\bibinfo  {journal} {Phys. Rev. B}\ }\textbf {\bibinfo {volume} {94}},\
  \bibinfo {pages} {165405} (\bibinfo {year} {2016})}\BibitemShut {NoStop}%
\bibitem [{\citenamefont {Andrade}\ \emph {et~al.}(2022)\citenamefont
  {Andrade}, \citenamefont {Carrillo-Bastos}, \citenamefont {Asmar},\ and\
  \citenamefont {Naumis}}]{Kekule_Naumis_PhysRevB.106.195413}%
  \BibitemOpen
  \bibfield  {author} {\bibinfo {author} {\bibfnamefont {E.}~\bibnamefont
  {Andrade}}, \bibinfo {author} {\bibfnamefont {R.}~\bibnamefont
  {Carrillo-Bastos}}, \bibinfo {author} {\bibfnamefont {M.~M.}\ \bibnamefont
  {Asmar}},\ and\ \bibinfo {author} {\bibfnamefont {G.~G.}\ \bibnamefont
  {Naumis}},\ }\href {https://doi.org/10.1103/PhysRevB.106.195413} {\bibfield
  {journal} {\bibinfo  {journal} {Phys. Rev. B}\ }\textbf {\bibinfo {volume}
  {106}},\ \bibinfo {pages} {195413} (\bibinfo {year} {2022})}\BibitemShut
  {NoStop}%
\bibitem [{\citenamefont {Iurov}\ \emph {et~al.}(2020)\citenamefont {Iurov},
  \citenamefont {Zhemchuzhna}, \citenamefont {Fekete}, \citenamefont {Gumbs},\
  and\ \citenamefont {Huang}}]{PhysRevResearch.2.043245}%
  \BibitemOpen
  \bibfield  {author} {\bibinfo {author} {\bibfnamefont {A.}~\bibnamefont
  {Iurov}}, \bibinfo {author} {\bibfnamefont {L.}~\bibnamefont {Zhemchuzhna}},
  \bibinfo {author} {\bibfnamefont {P.}~\bibnamefont {Fekete}}, \bibinfo
  {author} {\bibfnamefont {G.}~\bibnamefont {Gumbs}},\ and\ \bibinfo {author}
  {\bibfnamefont {D.}~\bibnamefont {Huang}},\ }\href
  {https://doi.org/10.1103/PhysRevResearch.2.043245} {\bibfield  {journal}
  {\bibinfo  {journal} {Phys. Rev. Res.}\ }\textbf {\bibinfo {volume} {2}},\
  \bibinfo {pages} {043245} (\bibinfo {year} {2020})}\BibitemShut {NoStop}%
\bibitem [{\citenamefont {B{\o}ggild}\ \emph {et~al.}(2017)\citenamefont
  {B{\o}ggild}, \citenamefont {Caridad}, \citenamefont {Stampfer},
  \citenamefont {Calogero}, \citenamefont {Papior},\ and\ \citenamefont
  {Brandbyge}}]{Boggild2017}%
  \BibitemOpen
  \bibfield  {author} {\bibinfo {author} {\bibfnamefont {P.}~\bibnamefont
  {B{\o}ggild}}, \bibinfo {author} {\bibfnamefont {J.~M.}\ \bibnamefont
  {Caridad}}, \bibinfo {author} {\bibfnamefont {C.}~\bibnamefont {Stampfer}},
  \bibinfo {author} {\bibfnamefont {G.}~\bibnamefont {Calogero}}, \bibinfo
  {author} {\bibfnamefont {N.~R.}\ \bibnamefont {Papior}},\ and\ \bibinfo
  {author} {\bibfnamefont {M.}~\bibnamefont {Brandbyge}},\ }\href
  {https://doi.org/10.1038/ncomms15783} {\bibfield  {journal} {\bibinfo
  {journal} {Nature Communications}\ }\textbf {\bibinfo {volume} {8}},\
  \bibinfo {pages} {15783} (\bibinfo {year} {2017})}\BibitemShut {NoStop}%
\bibitem [{\citenamefont {Chakraborti}\ \emph {et~al.}(2024)\citenamefont
  {Chakraborti}, \citenamefont {Gorini}, \citenamefont {Knothe}, \citenamefont
  {Liu}, \citenamefont {Makk}, \citenamefont {Parmentier}, \citenamefont
  {Perconte}, \citenamefont {Richter}, \citenamefont {Roulleau}, \citenamefont
  {Sacépé}, \citenamefont {Schönenberger},\ and\ \citenamefont
  {Yang}}]{Chakraborti_Makk_Peter_review_2024}%
  \BibitemOpen
  \bibfield  {author} {\bibinfo {author} {\bibfnamefont {H.}~\bibnamefont
  {Chakraborti}}, \bibinfo {author} {\bibfnamefont {C.}~\bibnamefont {Gorini}},
  \bibinfo {author} {\bibfnamefont {A.}~\bibnamefont {Knothe}}, \bibinfo
  {author} {\bibfnamefont {M.-H.}\ \bibnamefont {Liu}}, \bibinfo {author}
  {\bibfnamefont {P.}~\bibnamefont {Makk}}, \bibinfo {author} {\bibfnamefont
  {F.~D.}\ \bibnamefont {Parmentier}}, \bibinfo {author} {\bibfnamefont
  {D.}~\bibnamefont {Perconte}}, \bibinfo {author} {\bibfnamefont
  {K.}~\bibnamefont {Richter}}, \bibinfo {author} {\bibfnamefont
  {P.}~\bibnamefont {Roulleau}}, \bibinfo {author} {\bibfnamefont
  {B.}~\bibnamefont {Sacépé}}, \bibinfo {author} {\bibfnamefont
  {C.}~\bibnamefont {Schönenberger}},\ and\ \bibinfo {author} {\bibfnamefont
  {W.}~\bibnamefont {Yang}},\ }\href {https://doi.org/10.1088/1361-648X/ad46bc}
  {\bibfield  {journal} {\bibinfo  {journal} {Journal of Physics: Condensed
  Matter}\ }\textbf {\bibinfo {volume} {36}},\ \bibinfo {pages} {393001}
  (\bibinfo {year} {2024})}\BibitemShut {NoStop}%
\bibitem [{\citenamefont {Berry}\ and\ \citenamefont
  {Upstill}(1980)}]{Catastrophe_Optics_Berry_089}%
  \BibitemOpen
  \bibfield  {author} {\bibinfo {author} {\bibfnamefont {M.~V.}\ \bibnamefont
  {Berry}}\ and\ \bibinfo {author} {\bibfnamefont {C.}~\bibnamefont
  {Upstill}},\ }\bibinfo {title} {Catastrophe optics: morphologies of caustics
  and their diffraction patterns},\ in\ \href@noop {} {\emph {\bibinfo
  {booktitle} {Progress in Optics XVIII}}},\ \bibinfo {editor} {edited by\
  \bibinfo {editor} {\bibfnamefont {E.}~\bibnamefont {Wolf}}}\ (\bibinfo
  {publisher} {North-Holland},\ \bibinfo {year} {1980})\ pp.\ \bibinfo {pages}
  {257--346}\BibitemShut {NoStop}%
\bibitem [{\citenamefont {Berry}(1981)}]{sing_wave_caustics_Berry_105}%
  \BibitemOpen
  \bibfield  {author} {\bibinfo {author} {\bibfnamefont {M.~V.}\ \bibnamefont
  {Berry}},\ }\bibinfo {title} {Singularities in waves},\ in\ \href@noop {}
  {\emph {\bibinfo {booktitle} {Les Houches Lecture Series Session XXXV}}},\
  \bibinfo {editor} {edited by\ \bibinfo {editor} {\bibfnamefont
  {R.}~\bibnamefont {Balian}}, \bibinfo {editor} {\bibfnamefont
  {M.}~\bibnamefont {Kleman}},\ and\ \bibinfo {editor} {\bibfnamefont {J.-P.}\
  \bibnamefont {Poirier}}}\ (\bibinfo  {publisher} {Amsterdam: North-Holland},\
  \bibinfo {year} {1981})\ pp.\ \bibinfo {pages} {453--543}\BibitemShut
  {NoStop}%
\bibitem [{\citenamefont {Landau}\ \emph {et~al.}(1986)\citenamefont {Landau},
  \citenamefont {Pitaevskii}, \citenamefont {Lifshitz},\ and\ \citenamefont
  {Kosevich}}]{Landau1986_VII}%
  \BibitemOpen
  \bibfield  {author} {\bibinfo {author} {\bibfnamefont {L.~D.}\ \bibnamefont
  {Landau}}, \bibinfo {author} {\bibfnamefont {L.~P.}\ \bibnamefont
  {Pitaevskii}}, \bibinfo {author} {\bibfnamefont {E.~M.}\ \bibnamefont
  {Lifshitz}},\ and\ \bibinfo {author} {\bibfnamefont {A.~M.}\ \bibnamefont
  {Kosevich}},\ }\href
  {http://www.amazon.com/Theory-Elasticity-Third-Theoretical-Physics/dp/075062633X/ref=sr_1_16?ie=UTF8&s=books&qid=1280929419&sr=8-16}
  {\emph {\bibinfo {title} {Theory of Elasticity}}},\ \bibinfo {edition} {3rd}\
  ed.\ (\bibinfo  {publisher} {Butterworth-Heinemann},\ \bibinfo {year}
  {1986})\BibitemShut {NoStop}%
\bibitem [{\citenamefont {Stoker}(1989)}]{stoker1989differential:book}%
  \BibitemOpen
  \bibfield  {author} {\bibinfo {author} {\bibfnamefont {J.}~\bibnamefont
  {Stoker}},\ }\href@noop {} {\emph {\bibinfo {title} {Differential
  Geometry}}},\ Wiley Classics Library\ (\bibinfo  {publisher} {Wiley},\
  \bibinfo {year} {1989})\BibitemShut {NoStop}%
\bibitem [{saj()}]{sajat_suppl_Mat_multirefringence:cikk}%
  \BibitemOpen
  \href@noop {} {}\bibinfo {note} {See Supplemental Material at
  URL-will-be-inserted-by-publisher for details of calculation.}\BibitemShut
  {Stop}%
\bibitem [{\citenamefont {Guinea}\ \emph {et~al.}(2006)\citenamefont {Guinea},
  \citenamefont {Castro~Neto},\ and\ \citenamefont
  {Peres}}]{Guinea_stacks_PhysRevB.73.245426}%
  \BibitemOpen
  \bibfield  {author} {\bibinfo {author} {\bibfnamefont {F.}~\bibnamefont
  {Guinea}}, \bibinfo {author} {\bibfnamefont {A.~H.}\ \bibnamefont
  {Castro~Neto}},\ and\ \bibinfo {author} {\bibfnamefont {N.~M.~R.}\
  \bibnamefont {Peres}},\ }\href {https://doi.org/10.1103/PhysRevB.73.245426}
  {\bibfield  {journal} {\bibinfo  {journal} {Phys. Rev. B}\ }\textbf {\bibinfo
  {volume} {73}},\ \bibinfo {pages} {245426} (\bibinfo {year}
  {2006})}\BibitemShut {NoStop}%
\bibitem [{\citenamefont {Min}\ and\ \citenamefont
  {MacDonald}(2008)}]{Min_MacDonald_10.1143/PTPS.176.227}%
  \BibitemOpen
  \bibfield  {author} {\bibinfo {author} {\bibfnamefont {H.}~\bibnamefont
  {Min}}\ and\ \bibinfo {author} {\bibfnamefont {A.~H.}\ \bibnamefont
  {MacDonald}},\ }\href {https://doi.org/10.1143/PTPS.176.227} {\bibfield
  {journal} {\bibinfo  {journal} {Progress of Theoretical Physics Supplement}\
  }\textbf {\bibinfo {volume} {176}},\ \bibinfo {pages} {227} (\bibinfo {year}
  {2008})}\BibitemShut {NoStop}%
\bibitem [{\citenamefont {Yuan}\ \emph {et~al.}(2011)\citenamefont {Yuan},
  \citenamefont {Rold\'an},\ and\ \citenamefont
  {Katsnelson}}]{Katsnelson_ABC_PRB.84.125455}%
  \BibitemOpen
  \bibfield  {author} {\bibinfo {author} {\bibfnamefont {S.}~\bibnamefont
  {Yuan}}, \bibinfo {author} {\bibfnamefont {R.}~\bibnamefont {Rold\'an}},\
  and\ \bibinfo {author} {\bibfnamefont {M.~I.}\ \bibnamefont {Katsnelson}},\
  }\href {https://doi.org/10.1103/PhysRevB.84.125455} {\bibfield  {journal}
  {\bibinfo  {journal} {Phys. Rev. B}\ }\textbf {\bibinfo {volume} {84}},\
  \bibinfo {pages} {125455} (\bibinfo {year} {2011})}\BibitemShut {NoStop}%
\bibitem [{\citenamefont {Duppen}\ and\ \citenamefont
  {Peeters}(2013)}]{Van_Duppen_Peeters_2013}%
  \BibitemOpen
  \bibfield  {author} {\bibinfo {author} {\bibfnamefont {B.~V.}\ \bibnamefont
  {Duppen}}\ and\ \bibinfo {author} {\bibfnamefont {F.~M.}\ \bibnamefont
  {Peeters}},\ }\href {https://doi.org/10.1209/0295-5075/102/27001} {\bibfield
  {journal} {\bibinfo  {journal} {Europhysics Letters}\ }\textbf {\bibinfo
  {volume} {102}},\ \bibinfo {pages} {27001} (\bibinfo {year}
  {2013})}\BibitemShut {NoStop}%
\bibitem [{\citenamefont {Nakamura}\ and\ \citenamefont
  {Hirasawa}(2008)}]{PhysRevB.77.045429}%
  \BibitemOpen
  \bibfield  {author} {\bibinfo {author} {\bibfnamefont {M.}~\bibnamefont
  {Nakamura}}\ and\ \bibinfo {author} {\bibfnamefont {L.}~\bibnamefont
  {Hirasawa}},\ }\href {https://doi.org/10.1103/PhysRevB.77.045429} {\bibfield
  {journal} {\bibinfo  {journal} {Phys. Rev. B}\ }\textbf {\bibinfo {volume}
  {77}},\ \bibinfo {pages} {045429} (\bibinfo {year} {2008})}\BibitemShut
  {NoStop}%
\end{thebibliography}%


\begin{thebibliography}{10}%
\makeatletter
\providecommand \@ifxundefined [1]{%
 \@ifx{#1\undefined}
}%
\providecommand \@ifnum [1]{%
 \ifnum #1\expandafter \@firstoftwo
 \else \expandafter \@secondoftwo
 \fi
}%
\providecommand \@ifx [1]{%
 \ifx #1\expandafter \@firstoftwo
 \else \expandafter \@secondoftwo
 \fi
}%
\providecommand \natexlab [1]{#1}%
\providecommand \enquote  [1]{``#1''}%
\providecommand \bibnamefont  [1]{#1}%
\providecommand \bibfnamefont [1]{#1}%
\providecommand \citenamefont [1]{#1}%
\providecommand \href@noop [0]{\@secondoftwo}%
\providecommand \href [0]{\begingroup \@sanitize@url \@href}%
\providecommand \@href[1]{\@@startlink{#1}\@@href}%
\providecommand \@@href[1]{\endgroup#1\@@endlink}%
\providecommand \@sanitize@url [0]{\catcode `\\12\catcode `\$12\catcode
  `\&12\catcode `\#12\catcode `\^12\catcode `\_12\catcode `\%12\relax}%
\providecommand \@@startlink[1]{}%
\providecommand \@@endlink[0]{}%
\providecommand \url  [0]{\begingroup\@sanitize@url \@url }%
\providecommand \@url [1]{\endgroup\@href {#1}{\urlprefix }}%
\providecommand \urlprefix  [0]{URL }%
\providecommand \Eprint [0]{\href }%
\providecommand \doibase [0]{https://doi.org/}%
\providecommand \selectlanguage [0]{\@gobble}%
\providecommand \bibinfo  [0]{\@secondoftwo}%
\providecommand \bibfield  [0]{\@secondoftwo}%
\providecommand \translation [1]{[#1]}%
\providecommand \BibitemOpen [0]{}%
\providecommand \bibitemStop [0]{}%
\providecommand \bibitemNoStop [0]{.\EOS\space}%
\providecommand \EOS [0]{\spacefactor3000\relax}%
\providecommand \BibitemShut  [1]{\csname bibitem#1\endcsname}%
\let\auto@bib@innerbib\@empty
%</preamble>
\bibitem [{\citenamefont {Schwabl}(1992)}]{Schwabl_QM:book}%
  \BibitemOpen
  \bibfield  {author} {\bibinfo {author} {\bibfnamefont {F.}~\bibnamefont
  {Schwabl}},\ }\href {https://doi.org/10.1007/978-3-662-02703-5} {\emph
  {\bibinfo {title} {Quantum Mechanics}}}\ (\bibinfo  {publisher}
  {Springer-Verlag Berlin Heidelberg},\ \bibinfo {year} {1992})\BibitemShut
  {NoStop}%
\bibitem [{\citenamefont {Guinea}\ \emph {et~al.}(2006)\citenamefont {Guinea},
  \citenamefont {Castro~Neto},\ and\ \citenamefont
  {Peres}}]{Guinea_stacks_PhysRevB.73.245426}%
  \BibitemOpen
  \bibfield  {author} {\bibinfo {author} {\bibfnamefont {F.}~\bibnamefont
  {Guinea}}, \bibinfo {author} {\bibfnamefont {A.~H.}\ \bibnamefont
  {Castro~Neto}},\ and\ \bibinfo {author} {\bibfnamefont {N.~M.~R.}\
  \bibnamefont {Peres}},\ }\href {https://doi.org/10.1103/PhysRevB.73.245426}
  {\bibfield  {journal} {\bibinfo  {journal} {Phys. Rev. B}\ }\textbf {\bibinfo
  {volume} {73}},\ \bibinfo {pages} {245426} (\bibinfo {year}
  {2006})}\BibitemShut {NoStop}%
\bibitem [{\citenamefont {Min}\ and\ \citenamefont
  {MacDonald}(2008)}]{Min_MacDonald_10.1143/PTPS.176.227}%
  \BibitemOpen
  \bibfield  {author} {\bibinfo {author} {\bibfnamefont {H.}~\bibnamefont
  {Min}}\ and\ \bibinfo {author} {\bibfnamefont {A.~H.}\ \bibnamefont
  {MacDonald}},\ }\href {https://doi.org/10.1143/PTPS.176.227} {\bibfield
  {journal} {\bibinfo  {journal} {Progress of Theoretical Physics Supplement}\
  }\textbf {\bibinfo {volume} {176}},\ \bibinfo {pages} {227} (\bibinfo {year}
  {2008})}\BibitemShut {NoStop}%
\bibitem [{\citenamefont {Yuan}\ \emph {et~al.}(2011)\citenamefont {Yuan},
  \citenamefont {Rold\'an},\ and\ \citenamefont
  {Katsnelson}}]{Katsnelson_ABC_PRB.84.125455}%
  \BibitemOpen
  \bibfield  {author} {\bibinfo {author} {\bibfnamefont {S.}~\bibnamefont
  {Yuan}}, \bibinfo {author} {\bibfnamefont {R.}~\bibnamefont {Rold\'an}},\
  and\ \bibinfo {author} {\bibfnamefont {M.~I.}\ \bibnamefont {Katsnelson}},\
  }\href {https://doi.org/10.1103/PhysRevB.84.125455} {\bibfield  {journal}
  {\bibinfo  {journal} {Phys. Rev. B}\ }\textbf {\bibinfo {volume} {84}},\
  \bibinfo {pages} {125455} (\bibinfo {year} {2011})}\BibitemShut {NoStop}%
\bibitem [{\citenamefont {Duppen}\ and\ \citenamefont
  {Peeters}(2013)}]{Van_Duppen_Peeters_2013}%
  \BibitemOpen
  \bibfield  {author} {\bibinfo {author} {\bibfnamefont {B.~V.}\ \bibnamefont
  {Duppen}}\ and\ \bibinfo {author} {\bibfnamefont {F.~M.}\ \bibnamefont
  {Peeters}},\ }\href {https://doi.org/10.1209/0295-5075/102/27001} {\bibfield
  {journal} {\bibinfo  {journal} {Europhysics Letters}\ }\textbf {\bibinfo
  {volume} {102}},\ \bibinfo {pages} {27001} (\bibinfo {year}
  {2013})}\BibitemShut {NoStop}%
\bibitem [{\citenamefont {Nakamura}\ and\ \citenamefont
  {Hirasawa}(2008)}]{PhysRevB.77.045429}%
  \BibitemOpen
  \bibfield  {author} {\bibinfo {author} {\bibfnamefont {M.}~\bibnamefont
  {Nakamura}}\ and\ \bibinfo {author} {\bibfnamefont {L.}~\bibnamefont
  {Hirasawa}},\ }\href {https://doi.org/10.1103/PhysRevB.77.045429} {\bibfield
  {journal} {\bibinfo  {journal} {Phys. Rev. B}\ }\textbf {\bibinfo {volume}
  {77}},\ \bibinfo {pages} {045429} (\bibinfo {year} {2008})}\BibitemShut
  {NoStop}%
\bibitem [{\citenamefont {Mirzakhani}\ \emph {et~al.}(2017)\citenamefont
  {Mirzakhani}, \citenamefont {Zarenia}, \citenamefont {Vasilopoulos},\ and\
  \citenamefont {Peeters}}]{PhysRevB.95.155434}%
  \BibitemOpen
  \bibfield  {author} {\bibinfo {author} {\bibfnamefont {M.}~\bibnamefont
  {Mirzakhani}}, \bibinfo {author} {\bibfnamefont {M.}~\bibnamefont {Zarenia}},
  \bibinfo {author} {\bibfnamefont {P.}~\bibnamefont {Vasilopoulos}},\ and\
  \bibinfo {author} {\bibfnamefont {F.~M.}\ \bibnamefont {Peeters}},\ }\href
  {https://doi.org/10.1103/PhysRevB.95.155434} {\bibfield  {journal} {\bibinfo
  {journal} {Phys. Rev. B}\ }\textbf {\bibinfo {volume} {95}},\ \bibinfo
  {pages} {155434} (\bibinfo {year} {2017})}\BibitemShut {NoStop}%
\bibitem [{\citenamefont {Xiong}\ \emph {et~al.}(2017)\citenamefont {Xiong},
  \citenamefont {Jiang}, \citenamefont {Song},\ and\ \citenamefont
  {Duan}}]{Xiong_2017}%
  \BibitemOpen
  \bibfield  {author} {\bibinfo {author} {\bibfnamefont {H.}~\bibnamefont
  {Xiong}}, \bibinfo {author} {\bibfnamefont {W.}~\bibnamefont {Jiang}},
  \bibinfo {author} {\bibfnamefont {Y.}~\bibnamefont {Song}},\ and\ \bibinfo
  {author} {\bibfnamefont {L.}~\bibnamefont {Duan}},\ }\href
  {https://doi.org/10.1088/1361-648X/aa6aac} {\bibfield  {journal} {\bibinfo
  {journal} {Journal of Physics: Condensed Matter}\ }\textbf {\bibinfo {volume}
  {29}},\ \bibinfo {pages} {215002} (\bibinfo {year} {2017})}\BibitemShut
  {NoStop}%
\bibitem [{\citenamefont {Abramowitz}\ and\ \citenamefont
  {Stegun}(1972)}]{Abramowitz1:book}%
  \BibitemOpen
  \bibfield  {author} {\bibinfo {author} {\bibfnamefont {M.}~\bibnamefont
  {Abramowitz}}\ and\ \bibinfo {author} {\bibfnamefont {I.}~\bibnamefont
  {Stegun}},\ }\href
  {https://store.doverpublications.com/products/9780486612720} {\emph {\bibinfo
  {title} {Handbook of Mathematical Functions}}},\ \bibinfo {edition} {9th}\
  ed.\ (\bibinfo  {publisher} {Dover Publication Inc.},\ \bibinfo {address}
  {New York, NY},\ \bibinfo {year} {1972})\BibitemShut {NoStop}%
\bibitem [{\citenamefont {Gradshteyn}\ and\ \citenamefont
  {Ryzhik}(2007)}]{gradshteyn2007:book}%
  \BibitemOpen
  \bibfield  {author} {\bibinfo {author} {\bibfnamefont {I.~S.}\ \bibnamefont
  {Gradshteyn}}\ and\ \bibinfo {author} {\bibfnamefont {I.~M.}\ \bibnamefont
  {Ryzhik}},\ }\href
  {https://shop.elsevier.com/books/table-of-integrals-series-and-products/zwillinger/978-0-08-047111-2}
  {\emph {\bibinfo {title} {Table of integrals, series, and products}}},\
  \bibinfo {edition} {7th}\ ed.\ (\bibinfo  {publisher} {Elsevier/Academic
  Press, Amsterdam},\ \bibinfo {year} {2007})\BibitemShut {NoStop}%
\end{thebibliography}%

%\begin{thebibliography}{10}	
%\end{thebibliography}

\end{document}

\title{Supplemental Materials:  Snell's law in multirefringent systems}

\author{ J{\'o}zsef Cserti}
\author{Áron Holló}
\affiliation{Department of Physics of Complex Systems,
	ELTE E{\"o}tv{\"o}s Lor{\'a}nd University,
	H-1117 Budapest, P\'azm\'any P{\'e}ter s{\'e}t\'any 1/A, Hungary}

\author{László Oroszlány}
\affiliation{Department of Physics of Complex Systems,
	ELTE E{\"o}tv{\"o}s Lor{\'a}nd University,
	H-1117 Budapest, P\'azm\'any P{\'e}ter s{\'e}t\'any 1/A, Hungary\\
Wigner Research Centre for Physics, H-1525, Budapest, Hungary
}

%\begin{center}
%	\textbf{\large Supplemental Materials: 
%		Multirefringence and anomalous caustics in multi-band systems}
%\end{center}

\maketitle

%%%%%%%%%% Merge with supplemental materials %%%%%%%%%%
%%%%%%%%%% Prefix a "S" to all equations, figures, tables and reset the counter %%%%%%%%%%
\setcounter{equation}{0}
\setcounter{figure}{0}
\setcounter{table}{0}
\setcounter{page}{1}
%\makeatletter
\renewcommand{\theequation}{S\arabic{equation}}
\renewcommand{\thefigure}{S\arabic{figure}}
\renewcommand{\bibnumfmt}[1]{[S#1]}
\renewcommand{\citenumfont}[1]{S#1}
\setcounter{section}{0}
\renewcommand{\thesection}{S-\Roman{section}}

\section{Scattering problem for $N$ layer rhombohedral graphene}
\label{ABC_scatt:sec}

Here we present the calculations of the scattering of the incident plane wave for the rhombohedral stacking (ABC) of graphene multilayer. 
The scattering process of the electron can be calculated by the generalization
of the well-known partial wave method (see, e.g.~\cite{Schwabl_QM:book}). 
To this end, we take the Hamiltonian for $N$-layer ABC-stacked graphene multilayer around the $K$ point (see~\cite{Guinea_stacks_PhysRevB.73.245426,Min_MacDonald_10.1143/PTPS.176.227,Katsnelson_ABC_PRB.84.125455,Van_Duppen_Peeters_2013,PhysRevB.77.045429}) given by 
\begin{subequations}
	\label{ABC_Hamilton:eq}
	\begin{align}
		H_{N} &=
		\begin{bmatrix}
			H_\mathbf{p} & \Gamma & 0 & \cdots & 0 \\
			\Gamma^+ & H_\mathbf{p} & \Gamma & \cdots & 0 \\
			0 & \Gamma^+   &  H_\mathbf{p} & \cdots & 0 \\
			\vdots & \vdots  & \vdots & \ddots & \Gamma  \\
			0 & 0 & 0 & \Gamma^+ & 	H_\mathbf{p} \\
		\end{bmatrix} 
		+ V(\mathbf{r}) \, \mathbb{I}_{2N\times2N},
		\\[2ex]
		H_\mathbf{p} &= v_F
		\begin{bmatrix}
			0 & p_- \\[1ex] 
			p_+ & 0
		\end{bmatrix}, 
		\quad
		\Gamma =
		\begin{bmatrix}
			0 & 0 \\[1ex]
			\gamma_1 & 0
		\end{bmatrix}  \quad \mathrm{and} \quad 
		V(\mathbf{r}) =
		\begin{cases}
			V_<, & \text{if $|\mathbf{r}|<R$ }, \\
			V_>, & \text{otherwise},
		\end{cases}
	\end{align}
\end{subequations}
where $p_\pm = (p_x \pm i p_y)$ 
with the two-dimensional momentum operator 
$\mathbf{p} = (p_x,p_y)= -i \hbar \, \partial/\partial  \mathbf{r}$, and $v_F = 3 a t /2\hbar$ is the Fermi velocity of the
monolayer graphene, $t \approx 3$~eV is the hopping integral between the $p_z$ orbitals of two
adjacent carbon atoms and $a \approx 1.42 \textup{~\AA}$ is the carbon-carbon distance in graphene.
Here $V_{<}$ and $V_{>}$ are the constant gating potentials inside ($r<R$) and outside  ($r>R$) of the circular region of radius $R$, respectively.
The Hamiltonian is given in the $2N$ component basis $(\psi_{A_1},\psi_{B_1},\psi_{A_2},\psi_{B_2}, \dots,  \psi_{A_N},\psi_{B_N} )$, 
where $\psi_{A_i}$ ($\psi_{B_i}$) are the envelope functions associated with the probability amplitudes of the wave functions on sublattice A (B) of the
$i$th layer ($i = 1,2,\dots, N$).  
Finally, $\mathbb{I}_{2N\times2N}$ is a $2N\times2N$ unit matrix.
Here, we take into account only the nearest interlayer hopping, $\gamma_1 \approx 0.4$~eV.  
The effective Hamiltonian for $K^\prime$ is obtained by
exchanging  $p_+ $ and  $p_- $.

\begin{figure*}[h]
	\centering
		\includegraphics[scale=0.4]{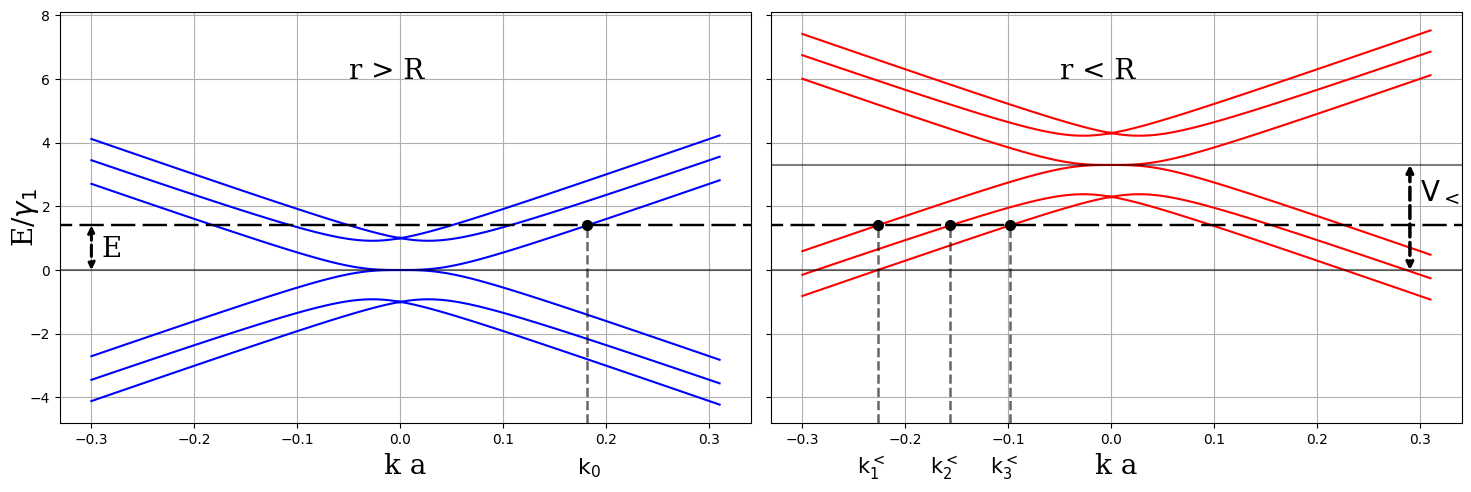}
	\caption{The energy dispersion $E(k)$ (in units of $\gamma_1$) as functions of the wave number $k$ 
  outside (left panel) and inside (right panel) of the circular region in the case of $N=3$ layers.
  We take the gate potentials $V_< = 3.3 \gamma_1$ and $V_> = 0$, and $R = 3000\, a$ in our numerical calculation. 
  The energy of the incoming electron is $E = 1.4 \gamma_1$ 
  from which the calculated wave numbers are $k_0 a= 0.181 $ for $r>R$, and 
 $k^{<}_1 a = -0.227, k^{<}_2 a =  -0.156, k^{<}_3 a= -0.098$ for $r<R$. 
 Thus, from Eq.~(3) of the main text %(\ref{n_def:eq}) 
 the refractive indices are
 $\hat{n} = \left\{-1.25, -0.86, -0.54 \right\}$.
		\label{disp_N_3:fig}}
\end{figure*}

Using the identities $\left[L_z, p_\pm\right] = \pm \hbar p_\pm $ 
(here $L_z = x p_y-y p_x $ is the $z$ component of the orbital angular momentum)
and the commutation relations between the Pauli matrices, 
it can be shown that the pseudo-angular momentum operator for 
$N$-layer ABC graphene
(for trilayer see Refs.~\cite{PhysRevB.95.155434,Xiong_2017}) 
\begin{align}
	\label{Jz_op:eq}
	J_z &= \mathbb{I}_{2N\times 2N} \otimes L_z
	+\frac{\hbar}{2}
	\begin{bmatrix}
		\sigma_0 & & \\
		& 3\sigma_0 & &  \\          
		%& & 5\sigma_0& \\    
		& &  \ddots \\
		& & & (2N-1)\sigma_0 \\
	\end{bmatrix}  -\frac{\hbar}{2}\, \mathbb{I}_{N\times N} \otimes \sigma_z,
\end{align}
commutes with the Hamiltonian, i.e., $\left[J_z,H_N\right] =0 $, where 
$\sigma_0$ is the $2\times 2$ unit matrix and $\sigma_z$ is the Pauli $z$ matrix.
Therefore, the pseudo-angular momentum is a conserved quantity in the scattering process. 
From now on, owing to the rotational symmetry of the Hamiltonian, it is convenient to use polar coordinates $(r, \varphi)$.
Thus, to construct the eigenfunctions of the Hamiltonian (\ref{ABC_Hamilton:eq}) 
we use the eigenfunction of the pseudo-angular momentum $J_z $, i.e.
$J_z\Psi_{k,m}(r,\varphi) = \hbar m \,	\Psi_{k,m}(r,\varphi)$ given by 
\begin{align}
	\label{Jz_eigen_vector:eq}
	\Psi_{k,m}(r,\varphi) &= \left( \begin{array}{c}
		C_1(k) \, Z_m(kr) \\
		C_2(k) \, Z_{m+1}(kr) \, e^{i\varphi} \\
		C_3(k) \, Z_{m+1}(kr) \, e^{i\varphi} \\
		C_4(k) \, Z_{m+2}(kr) \, e^{i2\varphi} \\
		C_5(k) \, Z_{m+2}(kr) \, e^{i2\varphi} \\
		C_6(k) \, Z_{m+3}(kr) \, e^{i3\varphi} \\
		C_7(k) \, Z_{m+3}(kr) \, e^{i3\varphi} \\
		\vdots \\
		C_{2N}(k) \, Z_{m+N}(kr) \, e^{iN\varphi}
	\end{array}  \right) \, e^{im\varphi},
\end{align}
where $Z_m(r)$ ) stands for the first-kind $J_m(r)$ and second-kind  $Y_m(r)$ Bessel functions, 
and for the Hankel functions of the first kind $H_m^{(1)}(r)$ and second kind $H_m^{(2)}(r)$ (here $m\in \mathbb{Z}$).  
For a given wave number $k$ the complex eigenvector $\mathbf{C}(k) = {\left[C_1(k),	C_2(k),\dots, C_{2N}(k)\right]}^T$ can be obtained from the Schr\"odinger equation 
$H_N \Psi_{k,m}(r,\varphi) = E(k) \, \Psi_{k,m}(r,\varphi)$ by 
substituting here the wave function (\ref{Jz_eigen_vector:eq}) resulting in the following eigenvalue equation for $\mathbf{C}(k)$: 
\begin{subequations}
	\label{Ham_separates:eq}
	\begin{align}
		\label{Ham_separates_a:eq}
		\left[\hbar  v_F k \, \Sigma + \mathcal{A} + V_{\lessgtr} \mathbb{I}_{2N\times2N}
		\right]\mathbf{C}(k) &= E(k) \, \mathbf{C}(k) , 
		\quad \mathrm{where} 
	\end{align}
	\begin{align}
		\Sigma &= \mathbb{I}_{N\times N}  \otimes  \sigma_y , \quad
		\mathcal{A} = \begin{pmatrix}
			0 & \Gamma & 0 & 0 & \cdots & 0 \\
			\Gamma^+ & 0 & \Gamma & 0 & \cdots & 0 \\
			0 & \Gamma^+ & 0 & \Gamma & \cdots & 0 \\
			\vdots & \vdots & \vdots & \vdots & \ddots & 0 \\
			0 & 0 & \cdots & \Gamma^+ & 0 & \Gamma \\
		\end{pmatrix}.
	\end{align}
\end{subequations}
To obtain Eq.~(\ref{Ham_separates:eq}) we used the following relation 
\begin{subequations}
	\begin{align}
		p_{\pm}Z_m(kr)e^{im\varphi} &=\pm i\hbar k Z_{m\pm1}(kr)e^{i(m\pm 1)\varphi},
	\end{align}
	where the momentum operator and the orbital angular momentum operator in polar coordinates are given by
	\begin{align} 
		p_\pm &= -i \hbar \, e^{\pm i \varphi}\, \left[\frac{\partial}{\partial r} \pm \frac{i}{r}\frac{\partial}{\partial \varphi}\right] 
		\quad \textrm{and} \quad  L_z = -i \hbar \, \partial_\varphi .  
	\end{align}
\end{subequations}
As an example, Fig.~\ref{disp_N_3:fig} shows the dispersion relations $E(k)$ inside ($r<R$) 
and outside ($r>R$) of the circular region. The energy $E$ of the incoming electron propagates from the left to the right direction with wave number $k_0$. 
The gating potential inside the circular region is a constant value of $V_<$ and zero outside. 
Inside the circular region, the allowed wave numbers with energy $E$ 
are $k^{<}_1, k^{<}_2$ and $k^{<}_3$. 
We set the value of $V_<$ in such a way that $\mathbf{v}(\mathbf{k}) \cdot \mathbf{k} < 0$ 
for the group velocities $\mathbf{v}(\mathbf{k})$ at the energy $E$, resulting in negative refraction indices.

In the scattering problem, the energy of the incident electron is given and we need to determine the allowed wave numbers corresponding to the propagating modes. 
However, from the eigenvalue equation~(\ref{Ham_separates:eq}), we obtain only the energy $E(k)$ as a function of wave number $k$. 
The inverse problem, namely, calculating the allowed wave number $k$ for a given energy $E$ is generally a difficult numerical problem involving the root finding. 
Fortunately, in our case, after rearranging 
Eq.~(\ref{Ham_separates_a:eq}) 
the wave number $k$ for a given energy $E$ can be determined from the eigenvalue equation of the matrix $\Lambda(E)$ as a convenient and standard procedure numerically:
\begin{subequations}
	\label{eigen_vec_C:eq}
	\begin{align}
		\Lambda(E) \mathbf{C}(k) &= 
		\hbar  v_F k \,\mathbf{C}(k),
		\quad \mathrm{where} \\[1ex]
		\Lambda(E) &=-\Sigma \left[\mathcal{A}+(V_{\lessgtr}-E)\mathbb{I}_{2N\times 2N}\right].
	\end{align}
\end{subequations}
Moreover, at the same time, we also obtained the $2N$ component vector $\mathbf{C}$ in the wave function given by Eq.~(\ref{Jz_eigen_vector:eq}).
Note that $\Sigma^{-1} = \Sigma$. 
In the above eigenvalue equation, there are $2N$ different numbers of 
the eigenvalues $k_\nu$ labeled by mode $\nu = 1,2,\dots,2N$. 
Note that some of the eigenvalues can be complex numbers. For the case of complex wave numbers, we should choose the one for which the partial cylindrical wave function tends to zero asymptotically.  

The scattering of the incident electron can be calculated by the 
generalization of the well-known partial wave method 
(see, e.g.,~\cite{Schwabl_QM:book}).
To this end, we construct the wave function inside and outside of the circular region as a linear combination of the eigenfunctions given by Eq.~(\ref{Jz_eigen_vector:eq}).
We assume that the wave number of the inward-propagating partial wave is $k_\nu$, corresponding to mode $\nu$. 
This is the $\nu$th solution of the equation $E(k_\nu) = E$, 
which is equivalent to the eigenvalues of Eq.~(\ref{eigen_vec_C:eq}). 
In the scattering problem, we choose $k_\nu$ as a real number. 
Then the wave function outside the circular region ($r > R$) can be written as 
\begin{subequations}
	\begin{align}
		\label{Psi_out_ansatz:eq}
		\Psi^{>}_{m,\nu}(r,\varphi) &=
		\mathbf{h}^{(2)}_{m,\nu} + \sum^{N}_{\beta = 1} S_{\nu \beta} \mathbf{h}^{(1)}_{m,\beta}, \quad \mathrm{for} \quad r > R,
	\end{align}
	where 
	\begin{align}
		\mathbf{h}^{(d)}(r,\varphi) = \left( \begin{array}{c}
			C_1^{>}(k_\nu) H^{(d)}_m(k_\nu r) \\
			C_2^{>}(k_\nu) H^{(d)}_{m+1}(k_\nu r)e^{i\varphi} \\
			C_3^{>}(k_\nu) H^{(d)}_{m+1}(k_\nu r)e^{i\varphi} \\
			C_4^{>}(k_\nu) H^{(d)}_{m+2}(k_\nu r)e^{i2\varphi} \\
			C_5^{>}(k_\nu) H^{(d)}_{m+2}(k_\nu r)e^{i2\varphi} \\
			C_6^{>}(k_\nu) H^{(d)}_{m+3}(k_\nu r)e^{i3\varphi} \\
			C_7^{>}(k_\nu) H^{(d)}_{m+3}(k_\nu r)e^{i3\varphi} \\
			\vdots \\
			C_{2N}^{>}(k_\nu) H^{(d)}_{m+N}(k_\nu r)e^{iN\varphi} \\
		\end{array}  \right)e^{im\varphi},
	\end{align}
\end{subequations}
and $\mathbf{h}^{(1)}$ and $\mathbf{h}^{(2)}$ are built up from  
the inward- and outward-propagating cylindrical waves, respectively, while  $\mathbf{C}^{>}(k_\nu) $ is the eigenvector in Eq.~(\ref{eigen_vec_C:eq}) with potential $V_>$,  and $S_{\nu\beta}$ is the scattering matrix of the partial waves from wave number $k_\nu$ to $k_\beta$. 
Note that $E(k_\nu) = E(k_\beta) = E$.
The scattering matrix $S_{\nu \beta}$ describes the scattering of 
a single incoming partial cylindrical wave $\mathbf{h}^{(1)}_{m, \beta}$ 
and can be obtained from the boundary condition of the wave function at $r=R$ (will be discussed below). 

The wave function inside the circular potential region ($0< r < R$)
should be built up from the partial waves (\ref{Jz_eigen_vector:eq}) with the  Bessel functions which is non-singular at $r=0$: 
\begin{subequations}
	\label{Psi_in_ansatz:eq}
	\begin{align}
		\Psi^{<}_{m,\nu}(r,\varphi) &=
		\sum^{N}_{\mu = 1} A_{\mu} \mathbf{J}_{m,\mu},
	\end{align}
	where 
	\begin{align}
		\mathbf{J}_{m, \mu}(r,\varphi) &= \left( \begin{array}{c}
			C_1^{<}(k_\mu) J_m(k_\mu r) \\
			C_2^{<}(k_\mu) J_{m+1}(k_\mu r)e^{i\varphi} \\
			C_3^{<}(k_\mu) J_{m+1}(k_\mu r)e^{i\varphi} \\
			C_4^{<}(k_\mu) J_{m+2}(k_\mu r)e^{i2\varphi} \\
			C_5^{<}(k_\mu) J_{m+2}(k_\mu r)e^{i2\varphi} \\
			C_6^{<}(k_\mu) J_{m+3}(k_\mu r)e^{i3\varphi} \\
			C_7^{<}(k_\mu) J_{m+3}(k_\mu r)e^{i3\varphi} \\
			\vdots \\
			C_{2N}^{<}(k_\mu) J_{m+N}(k_\mu r)e^{iN\varphi} \\
		\end{array}  \right)e^{im\varphi},
	\end{align}
\end{subequations}
where $\mathbf{C}^{<}(k_\mu) $ is the eigenvector in Eq.~(\ref{eigen_vec_C:eq}) with potential $V_<$ for wave number $k_\mu$ satisfying $E(k_\mu)=E$. 

Finally, the scattering matrix $S_{\nu \beta}$ and the amplitude $A_{\mu}$ are determined from the boundary conditions, i.e., the continuity of the total wave function: 
$\Psi^{>}_{m,\nu}(r = R, \varphi) = \Psi^{<}_{m,\nu}(r = R, \varphi)$, 
which leads to the following inhomogeneous linear equation system for a given $m$ and mode $\nu$:
\begin{align}
	\label{A_and_S_inhom:eq}
	\sum^{N}_{\mu = 1} A_{\mu} \mathbf{J}_{m,\mu} 
	&=  \mathbf{h}^{(2)}_{m,\nu} + \sum^{N}_{\beta = 1} S_{\nu \beta} \mathbf{h}^{(1)}_{m,\beta},
\end{align}
where the functions $\mathbf{J}_{m, \mu}, \mathbf{h}^{(1)}_{m, \nu}$ and $\mathbf{h}^{(2)}_{m, \nu}$ are evaluated at $r = R$. 
Note that the $\varphi$ dependence is dropped out from this equation. 
This system of equations has $2N$ number of equations with $N$ number of unknown amplitude $A_\nu$ and scattering matrix element $S_{\nu \beta}$. 

We now consider the scattering of an incident plane
wave propagating along the $x$ direction for $r > R$. 
Using Eq.~(\ref{Ham_separates:eq}) one can show that such an eigenstate 
$\Phi_\nu$ with a given energy $E$ and wave number $k_\nu$ is given by 
\begin{align}
	\Phi_\nu &=  e^{i k_\nu x} \, 
	\left( \begin{array}{c}
		C_1^{>}(k_\nu) \\
		i^{-1}\, C_2^{>}(k_\nu) \\
		i^{-1}\, C_3^{>}(k_\nu) \\
		i^{-2}\, C_4^{>}(k_\nu) \\
		i^{-2}\, C_5^{>}(k_\nu) \\
		i^{-3}\, C_6^{>}(k_\nu) \\
		i^{-3}\, C_7^{>}(k_\nu) \\
		\vdots \\
		i^{-N}\, C_{2N}^{>}(k_\nu) \\
	    \end{array} 
    \right),
\end{align}
where the vector $\mathbf{C}^{>}(k_\nu) $ is the eigenvector in Eq.~(\ref{eigen_vec_C:eq}) with potential $V_>$.
Then from the Jacobi--Anger expansion~\cite{Abramowitz1:book,gradshteyn2007:book}
%% Abramowitz - Stegun book ---> 9.1.41, page 366, using t = i exp(i \varphi)
%% Gradshteyn, I. S. and Ryzhik, I. M. book ---> 8.511,  1 & 4  page 982
\begin{align}
	e^{ik r \cos\varphi} = \sum_{m=-\infty}^{\infty}i^m J_m(kr)e^{im\varphi},
\end{align}
it can be shown that the plane wave $\Phi_\nu$ can be written as a
linear combination of incoming and outgoing cylindrical
waves:  
\begin{align}
	\Phi_\nu &= \sum_{m=-\infty}^{\infty} \frac{i^m}{2}
	\left(\mathbf{h}^{(1)}_{m,\nu} + \mathbf{h}^{(2)}_{m,\nu}\right) .
\end{align}

This expansion allows us to use the scattering matrix $S_{\nu \beta}$ and the amplitude $A_{\mu}$ determined from Eq.~(\ref{A_and_S_inhom:eq})
to derive the wave function describing the scattering of the plane wave. 
Then inside and outside of the circular region the wave functions 
$\Psi^{<}_\nu$ and $	\Psi^{>}_\nu$ become
\begin{subequations}
\begin{align}
	\Psi^{<}_\nu (r,\varphi) &= \sum_{m=-\infty}^{\infty} \sum_{\mu=1}^{N} \, 
	 \frac{i^m}{2} A_{\mu}\, \mathbf{J}_{m,\mu}, \quad \mathrm{for} \quad r<R, 
	 \\[1ex]
	 \label{scatt_Psi:eq}
	\Psi^{>}_\nu (r,\varphi) &= \Phi_\nu + 
	\sum_{m=-\infty}^{\infty} \sum_{\beta=1}^{N} \, 
	 \frac{i^m}{2}
	 \left( S_{\nu \beta}-\delta_{\nu \beta}\right)\, 
	 \mathbf{h}^{(1)}_{m,\beta},
	\quad \mathrm{for} \quad r>R.
\end{align}
\end{subequations}
The second term in (\ref{scatt_Psi:eq}) is the scattered wave due
to the scattering of the incident plane wave on the
scattering region described by the potential $V(\mathbf{r})$. 

\section{Birefringence in bilayer graphene and the emergence of anomalous caustics} 
\label{Results_N=2:sec}

\begin{figure}[!h]
	\centering
        \includegraphics[scale=0.4]{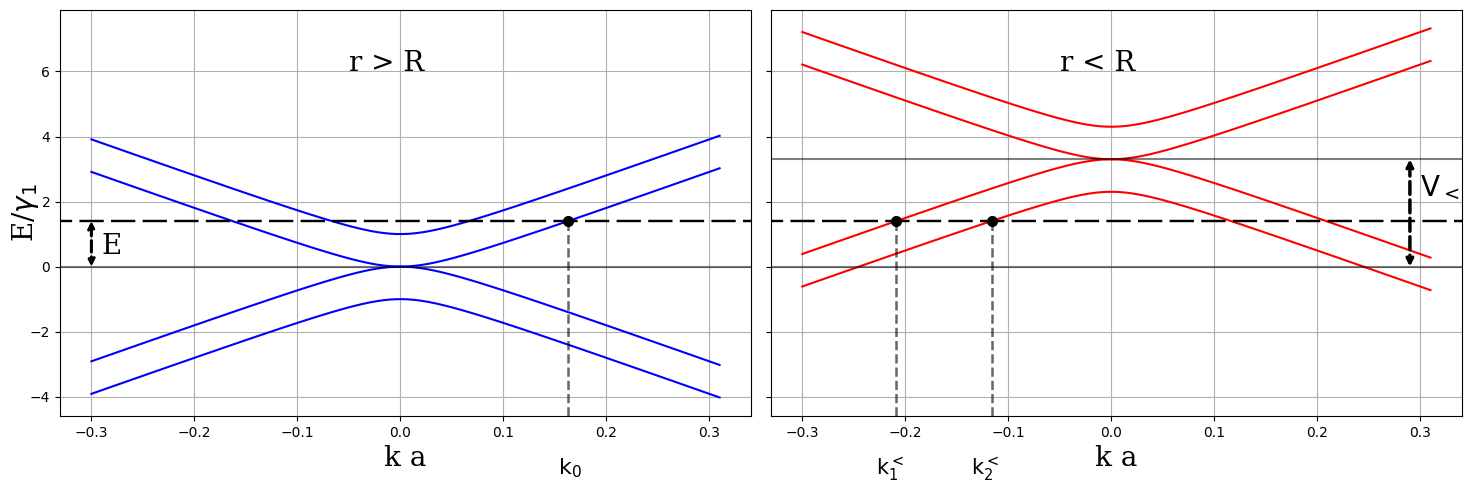}
	\caption{The energy dispersion $E(k)$ (in units of $\gamma_1$) as functions of the wave number $k$ 
  outside (left panel) and inside (right panel) of the circular region in the case of $N=2$ layers.
  We take the gate potentials $V_< = 3.3 \gamma_1$ and $V_> = 0$ in our numerical calculation. 
  The energy of the incoming electron is $E = 1.4 \gamma_1$ 
  from which the calculated wave numbers are  $k_0 a= 0.163 $ for $r>R$, and 
 $k^{<}_1 a = -0.209, k^{<}_2 a =  -0.116$ for $r<R$. 
 Thus, from Eq.~(3) of the main text %(\ref{n_def:eq}) 
 the refractive indices are 
 $\hat{n} = \left\{-1.28, -0.71 \right\}$.
		\label{disp_N_2:fig}}
\end{figure}

Figure~\ref{disp_N_2:fig} shows the dispersion relations $E(k)$ inside ($r<R$) 
and outside ($r>R$) of the circular region. The energy $E$ of the incoming electron propagates from the left to the right direction with wave number $k_0$. 
The gating potential inside the circular region is a constant value of $V_<$ and zero outside. 
Inside the circular region, the allowed wave numbers with energy $E$ 
are $k^{<}_1, k^{<}_2$ and $k^{<}_3$. 
We set the value of $V_<$ in such a way that $\mathbf{v}(\mathbf{k}) \cdot \mathbf{k} < 0$ 
for the group velocities $\mathbf{v}(\mathbf{k})$ at the energy $E$, resulting in negative refraction indices.

\begin{figure}[!h]
	\centering
         \includegraphics[scale=0.45]{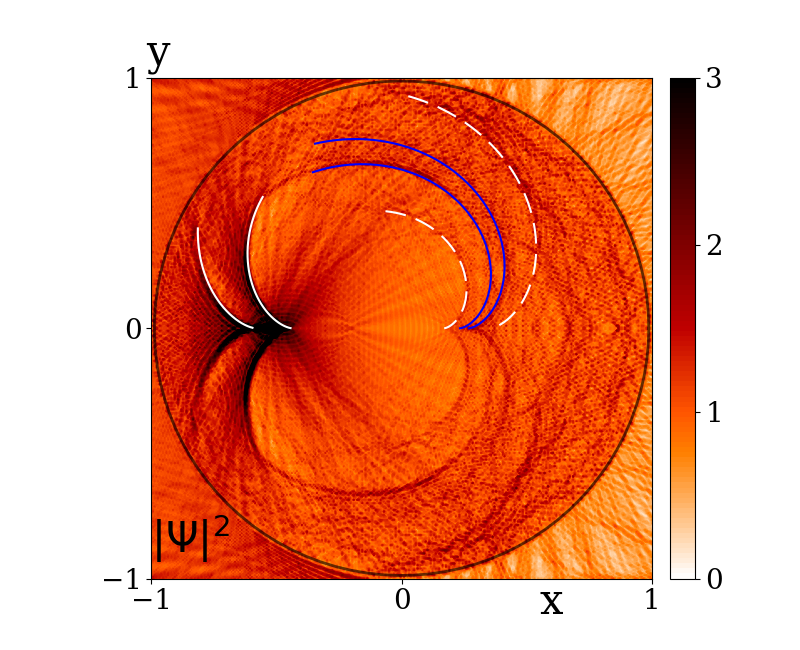}
	\caption{The intensity of the wave function inside the circular region 
		for ABC graphene with $N=2$ layers. 
  The intensity of the wave function inside the circularly gated potential for ABC graphene  (a) for $N=3$ layers (b) and $N=4$. 
		Using Eq.~(5) of the main text % ~(\ref{rc:eq}) 
  for $p=1$ and $p=2$ the regular (white and blue solid lines, respectively),  
            and for $p=2$ the anomalous (white dashed lines) caustic curves are plotted.
		The coordinates $x$ and $y$ are in units of $R$ and $R = 3000\, a$, where $a$ is the distance of two
adjacent carbon atoms.
	To have a better view, only half-sections of the symmetric caustic curves are shown. 
		\label{QM_N_2:fig}}
\end{figure}

Figure~\ref{QM_N_2:fig} shows the scattered wave function pattern inside the circular region for the case of $N=2$ layers. 
Using Eq.~(5) of the main text % (\ref{rc:eq}) 
we also plotted the regular caustics for $p=1$ and  $p=2$ chords (solid lines in Fig.~\ref{QM_N_2:fig}a), and the anomalous caustics for $p=2$ chords (dashed lines in Fig.~\ref{QM_N_2:fig}b).  
It can be seen from the figure that these curves agree well with the wave function pattern intensity maxima obtained from the full quantum mechanical scattering calculations. 

\begin{table}[htb]
	\centering
	\begin{tabular}{|c||c|c|} 
		\hline
		{$x^{(p=1)}_{\mathrm{cusp}}$}& $n_1 \rightarrow n_2$ & $x^{(p=2)}_{\mathrm{cusp}}$  \\ \hline
		-0.44 & \textcolor{red}{-0.71 $\rightarrow$ -1.28} & \textcolor{red}{0.17} \\ \hline
		-0.58 & -1.28 $\rightarrow$ -1.28 & 0.23 \\ \hline
		-  & -0.71 $\rightarrow$ -0.71 & 0.27 \\ \hline
            -  & \textcolor{red}{-1.28} $\rightarrow$ \textcolor{red}{-0.71} & \textcolor{red}{0.35} \\ \hline
	\end{tabular}
	\caption{The locations of the cusps $x^{(p=1)}_{\mathrm{cusp}}$ for $p=1$ chords, and $x^{(p=2)}_{\mathrm{cusp}}$
 for $p=2$ chords of the caustic curves shown in Fig.~\ref{QM_N_2:fig}  in case of $N=2$ layers.
 The locations are in units of $R$.
 For $p=1$ there are only regular caustic curves. 
 In the case of $p=2$ the two refractive indices $n_1$ and $n_2$ are the same for regular caustics, 
 while for different $n_1$ and $n_2$ the caustics are anomalous (red numbers).
 }
	\label{merged_cusps:table}
\end{table}

From Eq.~(3) of the main text %(\ref{n_def:eq}) 
with the allowed wave numbers for a given energy 
(see the caption of Fig.~\ref{QM_N_2:fig}) the two refractive indices of the circular regions are 
$\hat{n} = \left\{-1.28, -0.71 \right\}$. 
Then form Eq.~(6) of the main text %(\ref{cusp_anomal_p1:eq}) for regular caustics with $p=1$ and with $p=2$
we find $x^{(1)}_{\mathrm{cusp}} = -0.44 R$, and $x^{(1)}_{\mathrm{cusp}} = -0.58 R$, 
$x^{(2)}_{\mathrm{cusp}} = 0.23 R$, and $x^{(2)}_{\mathrm{cusp}} = 0.27 R$,
while for anomalous caustics form Eq.~(7) of the main text %(\ref{cusp_anomal_p2:eq}) with  $p=2$ 
we have $x^{(2)}_{\mathrm{cusp}} = 0.17 R$, and $x^{(2)}_{\mathrm{cusp}} = 0.35 R$.
In Fig.~\ref{QM_N_2:fig} one can see that the positions of these cusps
also agree very well with that obtained from the exact quantum calculations. 

%Our theory developed for multilayer rhombohedral graphene (see the Supplemental Material~\cite{sajat_suppl_Mat_multirefringence:cikk}) can be applied to bilayer systems ($N=2$). 
%The results presented in the Supplemental Material~\cite{sajat_suppl_Mat_multirefringence:cikk} agree with that obtained in Refs.~\onlinecite{bilayer_electron_flow_CsJ_PRB.80.075416}. 
%One can see from the wave function pattern, that the caustic curves obtained from~(\ref{rc:eq}) agree well with the location of intensity maxima of the calculated scattered wave function.
%Note that in contrast to the result in Ref.~\onlinecite{bilayer_electron_flow_CsJ_PRB.80.075416}, here the gate potential is set in such a way that for a given energy there are two distinct wave numbers (and thus two different refractive indices) in the circular region, therefore for $p=2$ there are two regular and two anomalous caustics.
\section{General equation for the caustics of straight lines } 
\label{gen_eq_caustics:sec}

Here we derive the equation for the caustics, i.e., the envelope of the family of straight lines in two dimensions. 
The equation of the family of straight lines depending on a parameter $\alpha$ can be given by 
\begin{align}
	\label{straight_line_1:eq}
	\left[\mathbf{r}-\mathbf{r}_0 (\alpha)\right]\cdot \mathbf{n} (\alpha) &= 0,
\end{align}
where $\mathbf{r}_0 (\alpha)$ is a point on the line and $\mathbf{n} (\alpha)$ is the normal vector to the line (not necessarily a unit vector) for a given parameter $\alpha$. 
%Note that both vectors depend on the parameter $\alpha$.
Now, differentiating Eq.~(\ref{straight_line_1:eq}) with respect to $\alpha$ 
we obtain
\begin{align}
	\label{straight_line_2:eq}
	\mathbf{r} \cdot \dot{\mathbf{n}}&= \dot{\mathbf{r}}_0\cdot \mathbf{n}+  \mathbf{r}_0 \cdot \dot{\mathbf{n}}.
\end{align}
Here the dot denotes the derivative with respect to $\alpha$.
The solution of Eqs.~(\ref{straight_line_1:eq}) and (\ref{straight_line_2:eq}) for $\mathbf{r}$ gives the parametric equation of the caustics curve $\mathbf{r}_c(\alpha)$. % ~\cite{stoker1989differential:book}.
To solve these two equations we are looking for the solution in the following form: 
\begin{align}
	\label{caustic_1:eq}
	\mathbf{r} &= \mathbf{r}_0+ C \, \mathbf{v}, 
\end{align}
where the vector $\mathbf{v}$ and the real number $C$ are to be determined.
Substituting this form into Eqs.~(\ref{straight_line_1:eq}) and (\ref{straight_line_2:eq}) we find
\begin{align}
	\label{caustic_2:eq}
	C \, \mathbf{v}\cdot \mathbf{n}  &= 0 
	\quad  \textrm{and} \quad 
	C\, \mathbf{v} \cdot \dot{\mathbf{n}} = \dot{\mathbf{r}}_0\cdot \mathbf{n}. 
\end{align}
For finite $C$ the first equation can be satisfied when $\mathbf{v}$ is perpendicular to $\mathbf{n}$, i.e., $\mathbf{v}\cdot \mathbf{n} = 0$.  
In two dimensions the vector $\mathbf{v}$ should be the direction vector of the straight line. 
While the second equation gives the solution for $C$, namely 
$C = \left(\dot{\mathbf{r}}_0\cdot \mathbf{n}\right)/ \left(\mathbf{v} \cdot \dot{\mathbf{n}}\right)$. 
Then, using Eq.~(\ref{caustic_1:eq})  the parametric equation of the caustics curve is 
\begin{align}
	\label{gen_eq_caustic:eq}
	\mathbf{r}_c (\alpha) \equiv \mathbf{r}    
	&= \mathbf{r}_0+ 
	\frac{\dot{\mathbf{r}}_0\cdot \mathbf{n}}{\mathbf{v} \cdot \dot{\mathbf{n}}}
	\,  \mathbf{v} ,
	%= \mathbf{r}_0+ 
	%\frac{\dot{\mathbf{r}}_0\cdot \mathbf{n}}{ \left| \dot{\mathbf{n}}\right|}
	%\,  \frac{\mathbf{v}}{\left|\mathbf{v}\right|} , 
\end{align}
where $\mathbf{r}_0$ is a given point on the straight line, $\mathbf{v}$ and $\mathbf{n}$ are the direction and the normal vector of the straight line, respectively. 
All vectors depend on the parameter  $\alpha$. 
Note that if $\mathbf{n}$ is normalized then in two dimensions $\mathbf{v} $ is parallel to $\dot{\mathbf{n}}$ thus it can be replaced by $\dot{\mathbf{n}}$  
in Eq.~(\ref{gen_eq_caustic:eq}). However, due to the normalization of $\mathbf{n}$, in general, this approach is less suitable for analytical calculations. 
Equation~(\ref{gen_eq_caustic:eq}) is the general equation for the caustic curve of the family of straight lines given by Eq.~(\ref{straight_line_1:eq}). 

\section{Application of the general equation for a family of chords in a circle} 
\label{chord_appl:sec}

Assume that the two endpoints of a chord in a circle with unit radius are given by 
\begin{align}
	\mathbf{r}_A &=
	\begin{bmatrix}
		\cos \varphi_A \\[1ex]
		\sin \varphi_A  
	\end{bmatrix},	\quad \mathbf{r}_B =
	\begin{bmatrix}
		\cos \varphi_B \\[1ex]
		\sin \varphi_B  
	\end{bmatrix},
\end{align}
where $\varphi_A$ and $\varphi_B$ depending on the parameter $\alpha$ are the polar angle of the vectors $\mathbf{r}_A$ and $\mathbf{r}_B$. 
The direction vector $\mathbf{v}$ and the normal vector $\mathbf{n}$ of the chords are  
\begin{align}
	\mathbf{v} &= \mathbf{r}_B -\mathbf{r}_A =
	\begin{bmatrix}
		\cos \varphi_B - \cos \varphi_A \\[1ex]
		\sin \varphi_B - \sin \varphi_A 
	\end{bmatrix},	\quad 
	\mathbf{n} =
	\begin{bmatrix}
		-\left(\sin \varphi_B - \sin \varphi_A\right)  \\[1ex]
		\cos \varphi_B - \cos \varphi_A  
	\end{bmatrix}.
\end{align}
Here $\mathbf{v}\cdot \mathbf{n} =0$.

The derivatives of $\mathbf{r}_A$ and $\mathbf{n}$ with respect to the parameter $\alpha$ are
\begin{align}
	\dot{\mathbf{r}}_A &=  
	\begin{bmatrix}
		-\sin \varphi_A \\[1ex]
		\cos \varphi_A 
	\end{bmatrix}\, \dot{\varphi}_A , \quad
	\dot{\mathbf{n}} = -
	\begin{bmatrix}
		\cos \varphi_B   \\[1ex]
		\sin \varphi_B   
	\end{bmatrix}\, \dot{\varphi}_B +
	\begin{bmatrix}
		\cos \varphi_A   \\[1ex]
		\sin \varphi_A   
	\end{bmatrix}\, \dot{\varphi}_A.
\end{align}

Using the general equation (\ref{gen_eq_caustic:eq}) for the caustics curve and with 
$\mathbf{r}_0 = \mathbf{r}_A$ we find 
\begin{align}
	\label{chord_caustic:eq}
	\mathbf{r}_c (\alpha)&= 
	\frac{\mathbf{r}_A\, \dot{\varphi}_B + 
		\mathbf{r}_B \dot{\varphi}_A}
	{\dot{\varphi}_A+\dot{\varphi}_B}. 
\end{align}

\bibliographystyle{apsrev4-2}
%\bibliography{/home/cserti/Dropbox/AD/BibTeX/cikkek.bib}
\bibliography{cikkek.bib}